\documentclass[aps,prd,10pt,superscriptaddress,twocolumn,nofootinbib,floatfix]{revtex4-2}

\usepackage[utf8]{inputenc}
\usepackage[paper=letterpaper,margin=1in]{geometry}

\usepackage{amsmath,amsfonts,amssymb}
\usepackage{mathrsfs}
\usepackage{mathtools}
\usepackage{amsthm}
\usepackage{bm}
\usepackage{bbm}
\usepackage{dsfont}
\usepackage{euscript}
\usepackage{upgreek}
\usepackage{braket}
\usepackage{tensor}
\usepackage{leftidx}
\usepackage{slashed}
\usepackage{cancel}
\allowdisplaybreaks[4]

\usepackage{starfont}
\usepackage{wasysym}

\usepackage{booktabs}
\newcommand{\eqstrut}{\rule[-3.2ex]{0pt}{8ex}}

\usepackage[dvipsnames]{xcolor}
\usepackage{color,soul}
\usepackage[normalem]{ulem}
\definecolor{outerspace}{rgb}{0.25, 0.29, 0.3}
\definecolor{scarlet}{rgb}{1.0, 0.13, 0.0}
\definecolor{princetonorange}{rgb}{1.0, 0.56, 0.0}
\definecolor{WildStrawberry}{rgb}{1.0, 0.26, 0.64}
\definecolor{rossocorsa}{rgb}{0.83, 0.0, 0.0}
\definecolor{navyblue}{rgb}{0.0, 0.0, 0.5}

\usepackage{graphicx}

\usepackage{multirow}
\usepackage{hhline}

\usepackage{changepage}
\usepackage{enumitem}
\usepackage{comment}

\newcommand{\ve}{\varepsilon}           
\newcommand{\ie}{\textit{i.e., }}       
\newcommand{\dd}{\mathrm{d}}            
\newcommand{\m}{\mathsf{m}}             
\newcommand{\p}{\mathsf{p}}             
\newcommand{\M}{\mathsf{M}}             
\newcommand{\GN}{G_{\rm N}}             
\newcommand{\be}{\begin{equation}}      
\newcommand{\ee}{\end{equation}}        
\newcommand{\defeq}{                    
  \mathrel{
    \mskip3mu
    \ooalign{
      \hfil=\hfil\cr
      \raise0.68ex\hbox to 0pt{\hskip -0.35ex\hss.\hss}\hidewidth\cr
      \raise+0.23ex\hbox to 0pt{\hskip -0.35ex\hss.\hss}\hidewidth\cr
}}}
\usepackage{scalerel}

\newlength{\halftextheight}
\newcommand\reallywidehat[1]{\arraycolsep=0pt\relax
\begin{array}{c}
\stretchto{
  \scaleto{
    \scalerel*[\widthof{\ensuremath{#1}}]{\kern-.5pt\bigwedge\kern-.5pt}
    {\rule[-\halftextheight]{1ex}{\textheight}}
  }{\textheight}
}{0.5ex}\\
#1\\
\rule{-1ex}{0ex}
\end{array}
}
\makeatletter
\newcommand{\dal}{\mathop{\mathpalette\dal@\relax}}
\newcommand{\dal@}[2]{
  \begingroup
  \sbox\z@{$\m@th#1\square$}
  \dimen0=\fontdimen8
    \ifx#1\displaystyle\textfont\else
    \ifx#1\textstyle\textfont\else
    \ifx#1\scriptstyle\scriptfont\else
    \scriptscriptfont\fi\fi\fi3
  \makebox[\wd\z@]{
    \hbox to \ht\z@{
      \vrule width \dimen0
      \kern-\dimen0
      \vbox to \ht\z@{
        \hrule height \dimen0 width \ht\z@
        \vss
        \hrule height 2\dimen0
      }
      \kern-2.5\dimen0
      \vrule width 2.5\dimen0
    }
  }
  \endgroup
}
\makeatother

\usepackage{hyperref}
\hypersetup{
    colorlinks,
    citecolor=WildStrawberry,
    filecolor=WildStrawberry,
    linkcolor=MidnightBlue,
    urlcolor=MidnightBlue
}
\usepackage{cleveref}

\crefname{section}{Sec.}{Secs.}
\crefname{equation}{Eq.}{Eqs.}
\crefname{figure}{Fig.}{Figs.}
\crefname{table}{Table}{Tables}
\crefname{appendix}{Appendix}{Appendices}
\crefname{defi}{Definition}{Definitions}
\crefname{theorem}{Theorem}{Theorems}
\crefname{proposition}{Proposition}{Propositions}
\crefname{corollary}{Corollary}{Corollaries}
\crefname{lemma}{Lemma}{Lemmas}
\crefname{conjecture}{Conjecture}{Conjectures}

\begin{document}

\title{Thin shells in Lovelock and Quasi-topological gravities}

\author{Pablo Bueno}
\email{pablobueno@ub.edu}
\affiliation{Departament de F\'isica Qu\`antica i Astrof\'isica, Institut de Ci\`encies del Cosmos\\
 Universitat de Barcelona, Mart\'i i Franqu\`es 1, E-08028 Barcelona, Spain}

\author{Pablo A. Cano}
\email{pablocano@um.es}
\affiliation{Departamento de F\'isica, Universidad de Murcia, Campus de Espinardo, 30100 Murcia, Spain}

\author{Robie A. Hennigar}
\email{robie.a.hennigar@durham.ac.uk}
\affiliation{Centre for Particle Theory, Department of Mathematical Sciences,\\ Durham University, Durham DH1 3LE, UK}

\author{\'Angel J. Murcia}
\email{ajm386@cam.ac.uk}
\affiliation{Department of Applied Mathematics and Theoretical Physics, University of Cambridge, Wilberforce Road, Cambridge CB3 0WA, United Kingdom}
    
\author{Aitor Vicente-Cano}
\email{avicentecano@icc.ub.edu}
\affiliation{Departament de F\'isica Qu\`antica i Astrof\'isica, Institut de Ci\`encies del Cosmos\\
 Universitat de Barcelona, Mart\'i i Franqu\`es 1, E-08028 Barcelona, Spain}


\begin{abstract}
    We derive the junction conditions for spherically symmetric thin shells in Lovelock and Quasi-topological gravities along two independent routes: from the distributional form of the field equations, and from the boundary term that makes the variational problem well-posed in the non-null case. The former method is ambiguous as it stands, since the equations contain products of Heaviside step functions with a Dirac delta. We show that the singular parts of the field equations are exact differentials in the jumping variables, so that the Vol’pert chain rule for functions of bounded variation fixes these singular combinations uniquely and no products of two deltas survive for a single shell, finding that both routes then agree. The same rule yields the null junction conditions, for which no boundary term is available in Quasi-topological gravities, together with the mass conservation law of the shell. These results clarify recent questions concerning the existence of thin shells in these theories.  In particular, we show that previous claims that the junction conditions force the shell mass to vary with radius are incorrect and arise from treating a trapezoidal approximation to the relevant distributional products as exact. Our results also provide a basis for the recent analyses of mass inflation in these theories.
\end{abstract}

\onecolumngrid

\maketitle

\twocolumngrid

\vspace{-.3cm}

\section{Introduction}

The matching of two spacetimes across a hypersurface has been a subject of longstanding interest. Junction conditions in general relativity (GR) were first studied by Lanczos~\cite{Lanczos1922,Lanczos1924}, and later given a geometric formulation by Darmois~\cite{Darmois1927}. The modern treatment is due to Israel~\cite{Israel:1966rt}, who developed their coordinate-independent form. Barrab\`es and Israel~\cite{Barrabes:1991ng} later extended the formalism to null shells, where the induced metric degenerates (see also~\cite{CJSClarke_1987}), while shells of arbitrary or changing causal character were considered in~\cite{Mars:1993mj,Senovilla:2018hrw}.

Beyond GR this procedure is no longer straightforward. The field equations may contain products of first and second derivatives, so the singular terms come multiplied by quantities that are discontinuous, and products of the form $\Theta\delta$ (where $\Theta$ is the Heaviside step function and $\delta$ the
Dirac delta) appear. This type of quantities is ill-defined in standard distribution theory. Ways to circumvent this issue have long been explored. One method is to replace the distributional derivation by a variational one, obtaining the junction conditions from the boundary term that makes the variational problem well-posed. This was done for Gauss--Bonnet in~\cite{Davis:2002gn}, for Lovelock in~\cite{Gravanis:2004kx}, and for Horndeski in~\cite{Padilla:2012ze}. More recently, the same approach has been applied to a family of quadratic theories~\cite{Ramirez:2024czb}. There are also alternatives that keep the distributional setting and impose the additional continuity conditions that remove these products altogether~\cite{Senovilla:2013vra,Senovilla:2014kua,Senovilla:2014yea,Reina:2015gxa,Eiroa:2017zwh,Senovilla:2018hrw}. In some cases, this allows for gravitational double layers, involving $\delta'$~\cite{Senovilla:2014kua}. Junction conditions for general gravitational theories have recently been addressed in~\cite{Senovilla:2026fby}. Another option is to replace linear distribution theory with algebras of generalized functions that permit the computation of non-linear products~\cite{Steinbauer:2006qi}. While this has been explored in~\cite{Chu:2021uec}, the approach still suffers from indeterminacies, leading us to stay within the linear setting.

Here our focus is the spherically symmetric (SS) sector of Quasi-topological  (QT) gravities~\cite{Oliva:2010eb,Myers:2010ru,Dehghani:2011vu,Ahmed:2017jod,Cisterna:2017umf,Bueno:2019ycr,Bueno:2022res,Moreno:2023rfl,Moreno:2023arp,Bueno:2026oyg}. QT gravities are a broad class of higher-curvature theories beyond GR that generalize Lovelock gravities~\cite{Lovelock1}. The defining feature of these theories is that they yield second-order equations for SS metrics. Consequently, their dynamics in spherical symmetry are captured by a class of two-dimensional Horndeski gravities upon dimensional reduction~\cite{Bueno:2024eig,Bueno:2024zsx}. Like Lovelock gravity, QT gravities satisfy a Birkhoff theorem in vacuum, ensuring the general SS solution is also static, which reduces the equations of motion to algebraic ones~\cite{Oliva:2010eb,Oliva:2011xu,Cisterna:2017umf,Bueno:2025qjk}. In contrast to Lovelock gravity, QT theories exist at every order in curvature in any dimension $D \ge 5$ except for $n=D/2$ in even dimensions.\footnote{It is also possible to define non-polynomial generalizations of QT theories which exist in $D = 4$ and also admit a two-dimensional Horndeski formulation in spherical symmetry~\cite{Bueno:2025zaj,Borissova:2026wmn,Colleaux:2019ckh,Colleaux:2017ibe,Colleaux:2026hat}. In fact, any two-dimensional Horndeski theory can arise as the spherical reduction of a higher-dimensional (non)-polynomial gravity~\cite{Carballo-Rubio:2025ntd,Borissova:2026krh}.}

QT gravities have recently attracted considerable attention after it was realized~\cite{Bueno:2024dgm} that by summing up infinite towers of these corrections, the vacuum solutions are regular black holes (RBHs) under very general conditions~\cite{Bueno:2024dgm,Hennigar:2025yqm,Fernandes:2025fnz,Bueno:2026dln}. As such, QT gravity has been explored as a framework in which questions about RBHs concerning \textit{dynamics}\footnote{Nevertheless, it has also proven to be useful a purely phenomenological approach to RBHs, with no definite dynamics, particularly for the study of the evaporation process or the mass inflation problem, see \emph{e.g.},~\cite{Carballo-Rubio:2018pmi,Carballo-Rubio:2022kad,Carballo-Rubio:2025fnc,Carballo-Rubio:2026gwg}.} can be tackled~\cite{Bueno:2025gjg,Bueno:2025tli,Bueno:2026dln,Frolov:2024hhe,Frolov:2026rcm,PinedoSoto:2026hfm,Frolov:2026tft,Borissova:2026prd,Borissova:2026rbi,Konoplya:2024kih,Konoplya:2024hfg,DiFilippo:2024mwm,Ma:2024olw,Cisterna:2024ksz,Ditta:2024iky,Wang:2024zlq,Hennigar:2025yqm,Fernandes:2025eoc,Fernandes:2025fnz,Cisterna:2025vxk,Ling:2025ncw,Eichhorn:2025pgy,Fernandes:2025mic,Aguayo:2025xfi,Boyanov:2025pes,Konoplya:2025uta,Hao:2025utc,Tan:2025hht,Frolov:2025ddw,Chen:2025iuy,Arbelaez:2026eaz,Frolov:2026byy,Li:2026mam,Tsuda:2026xjc,Konoplya:2026gim,Fan:2026mdu,Bueno:2026bby}. Several of these works have employed the junction conditions, which were obtained for non-null shells using the variational method in~\cite{Bueno:2024zsx}. Dynamical collapse of thin shells was explored in~\cite{Bueno:2024eig, Bueno:2024zsx, Bueno:2025zaj}, showing that the vacuum RBH solutions are indeed formed in collapse. Oppenheimer--Snyder collapse, for which the junction conditions are necessary at the surface of the collapsing star, was studied in~\cite{Bueno:2025gjg}. More recently, intersecting null shells were studied in~\cite{Frolov:2026rcm} to investigate the question of mass inflation for RBHs. While the junction conditions for null shells had not yet been developed in QT gravity, Ref.~\cite{Frolov:2026rcm} used only the DTR relation (also called DHBI relation) $f_A f_B = f_C f_D$~\cite{Dray:1984ha,Redmount:1985aoy}, which in GR follows from Birkhoff's theorem, the kinematical properties of the shell crossing geometry, and, of course, the underlying assumption that thin null shells can exist. Remarkably, the authors of~\cite{Frolov:2026rcm} find that if the fundamental QT length scale is of the order of the Planck length, mass inflation in macroscopic RBHs is insignificant. This holds true unless shell intersection occurs at a distance from the inner horizon much smaller than the Planck length, a regime where the classical description is altogether unreliable.

The recent work~\cite{DiFilippo:2026jpv} set out to address the gap in~\cite{Frolov:2026rcm} by identifying the junction conditions for SS null thin shells in QT gravities, and to do so within the distributional approach of Barrab\`es and Israel. The nonlinear field equations introduce distributional products such as $\Theta \delta$, which require extra care. Ref.~\cite{DiFilippo:2026jpv} explored several prescriptions and found that, apart from certain special cases, the shell mass depends on radius. This led the authors to question the existence of thin null shells in Lovelock and QT theories, with related concerns about non-null shells expressed in their Appendix A. These findings raise a question about their relation to earlier non-null junction conditions obtained from the variational principle~\cite{Gravanis:2004kx, Miskovic:2007mg, Guilleminot:2022psn} and distributional methods~\cite{Charmousis:2005ey, Chu:2021uec}. Since these approaches use different frameworks or assumptions, understanding the origin of the different conclusions requires a closer examination of the singular terms. 

In this work we address these questions by studying the dynamics of non-null and null thin shells, deriving the generalized Israel and the generalized Barrab\`es--Israel junction conditions, respectively. In the non-null case we make use of both standard methods: a distributional derivation from the field equations and a variational derivation based on boundary terms. In the former approach we shall encounter terms of the form $\Theta\delta$, products which are not defined in ordinary distribution theory. However, crucially, the ambiguous combinations appearing in the QT equations are exact differentials in the jumping variables. For example, if $\psi$ is the quantity which has a jump, the problematic products always arise in the form $h'(\psi) \partial \psi$ for some smooth function $h$. This fact allows us to fix them uniquely by employing the Vol'pert chain rule for functions of bounded variation~\cite{Volpert_1967}, which we review. Subsequently, we also derive the null junction conditions in the distributional sense, clarifying the origin of the different conclusions in Ref.~\cite{DiFilippo:2026jpv} and showing that null thin shells exist. This removes the obstruction to their use in~\cite{Frolov:2026rcm}, while a more detailed assessment of multi-shell crossing dynamics remains for future work. A bird's eye view of our results can be found in Table~\ref{table:results}, where we present the dynamical equations of null and non-null shells for general Lovelock and QT gravities, together with their trapezoidal approximations, which reproduce the corresponding expressions in Ref.~\cite{DiFilippo:2026jpv}.

The paper is organized as follows.
In Sec.~\ref{sec:qt} we present the QT theories together with their two-dimensional reduction.
In Sec.~\ref{sec:vaidya-warmup}, we study a Vaidya solution describing a black hole absorbing a smooth, compactly supported pulse of radiation in QT gravity. Examining the limit where the pulse becomes distributional allows for the key ideas required for the junction conditions to be introduced in a pedagogical manner. In Sec.~\ref{sec:nonnull} we then derive the generalized Israel junction conditions for non-null shells described as a perfect fluid, from both the distributional and variational approaches. In the process, we present the Vol'pert average as a way to solve the ambiguity of the products $\Theta\delta$. 
Analogously, in Sec.~\ref{sec:null} we find the generalized Barrab\`es--Israel junction conditions for null shells, also described as a perfect fluid, from the distributional perspective, using the Vol'pert average. 
Finally, in Sec.~\ref{sec:discussion}, we discuss the implications of our results for the existence of thin shells and outline future lines of research.
Additionally, in Appendix~\ref{app:resummation} we rederive the shell-equations by resumming the series with the rule $\Theta^{n-1}\delta = \delta/n$, while Appendix~\ref{app:general-junctions} provides a derivation of the junction conditions for general two-dimensional Horndeski theories (including members beyond the QT class).

\section{Lovelock and Quasi-topological gravities} \label{sec:qt}

A general higher-curvature theory built from contractions of the metric and the Riemann tensor, minimally coupled to matter, can be written as
\begin{equation}\label{eq:QTaction}
    I\!=\!\int \! \dd^D x \sqrt{|g|} \!\left(\frac{\mathcal{L}_{\rm QT}\big(R_{abcd},g^{ef}\big)}{16\pi \GN} + \mathcal{L}_{\rm matter}\right) \,,
\end{equation}
whose field equations are $\mathcal{E}_{ab} = 8\pi\GN T_{ab}$, with the generalized Einstein tensor~\cite{Padmanabhan:2011ex}
\begin{equation}\label{eq:Eab}
    \mathcal{E}_{ab}\equiv P_a{}^{cde}R_{bcde}-\frac{g_{ab}\mathcal{L}_{\rm QT}}{2} -2\nabla^c\nabla^d P_{acdb} \,,
\end{equation}
where
\begin{equation}
    P^{abcd}\equiv\frac{\partial\mathcal{L}_{\rm QT}}{\partial R_{abcd}} \,,
\end{equation}
and, in general, the field equations are fourth-order in derivatives.

We focus on QT gravities, a class of higher-curvature theories whose field equations become of second order when evaluated on spherically symmetric backgrounds~\cite{Oliva:2010eb,Myers:2010ru,Dehghani:2011vu,Ahmed:2017jod,Cisterna:2017umf,Bueno:2019ycr,Bueno:2022res,Moreno:2023rfl}. These theories generically satisfy a Birkhoff theorem, so their spherically symmetric vacuum solutions are static and unique~\cite{Bueno:2025qjk}. Their Lagrangian takes the form
\begin{equation}\label{genAc}
    \mathcal{L}_{\rm QT}\big(R_{abcd},g^{ef}\big) = R + \sum_{n\geq 2} \widetilde\alpha_n \,\mathcal{Z}_n \,,
\end{equation}
where $\widetilde\alpha_n$ are couplings of dimension length$^{2(n-1)}$. Non-trivial QT densities $\mathcal{Z}_n$ exist at every curvature order in $D\geq5$, except for $n=D/2$ in even $D$, and, remarkably, they can be generated recursively from the first five~\cite{Bueno:2019ycr}. Crucially, once evaluated on a spherically symmetric background, the whole tower collapses onto the two-dimensional description we set up next.

Before doing so, let us note that when the sum~\eqref{genAc} truncates at some $n_{\rm max}\leq \lfloor\frac{D-1}{2}\rfloor$, the spherical sector of that Lagrangian coincides with that of a Lovelock theory of the same order. For our purposes we leave $n_{\rm max}$ arbitrary, so that our results apply equally to Lovelock and QT theories, including the infinite towers that yield regular black holes.

\subsection{Effective two-dimensional description}
\label{sec:2d}

Let us write a general $D$-dimensional spherically symmetric metric in the form
\begin{equation}\label{eq:sphericmetric}
    \dd s^2 = \gamma_{\mu\nu}\,\dd x^{\mu}\dd x^{\nu} + \varphi(x^\rho)^2\,\dd\Omega^2_{D-2} \,,
\end{equation}
with $\gamma_{\mu\nu}$ a two-dimensional Lorentzian metric --- we will use Greek indices to refer to components of this two-dimensional submanifold --- and $\varphi$ a covariant areal radius. Evaluating~\eqref{eq:QTaction} on~\eqref{eq:sphericmetric} gives a two-dimensional ``dilaton-gravity'' model, and since the theory is second order on such configurations by construction, that model is a two-dimensional Horndeski theory~\cite{Horndeski:1974wa},
\begin{equation}\label{eq:2daction}
    I_{\rm 2d} = \frac{(D-2)\Omega_{D-2}}{16\pi \GN} \int \dd^{2} x \sqrt{|\gamma|}\, \mathcal{L}_{\rm 2d} (\gamma_{\mu\nu},\varphi) \,,
\end{equation}
where $\Omega_{D-2} = 2\pi^{(D-1)/2}/\Gamma\left(\tfrac{D-1}{2}\right)$ is the volume of the $(D-2)$-sphere, and the Lagrangian corresponding to a general $D$-dimensional QT gravity reads~\cite{Bueno:2024zsx}
\begin{align}\notag 
    &\mathcal{L}_{\rm 2d} = G_{2}(\varphi, X)-\Box\varphi G_{3}(\varphi, X)+G_{4}(\varphi, X)R\\  \label{eq:horn}&-2G_{4,X}(\varphi, X)\left((\Box\varphi)^2-\nabla_{\mu}\nabla_{\nu}\varphi\nabla^{\mu}\nabla^{\nu}\varphi\right) \,,
\end{align}
where, from now on, $R$, $\Box$ and $\nabla_\mu$ refer to $\gamma_{\mu\nu}$, comma denotes a partial derivative, and the $G_w$ depend on $\varphi$ and on
\begin{equation}\label{eq:psiXdef}
    X \equiv \nabla_{\mu}\varphi\nabla^{\mu}\varphi \,, \qquad \psi \equiv \frac{1-X}{\varphi^2} \,.
\end{equation}
The theory dependence is carried by the following \emph{characteristic function},
\begin{equation}\label{eq:hpsi}
    h(\psi)\equiv \psi + \sum_{n\geq 2} \alpha_n\, \psi^n \,, \qquad   \alpha_n \equiv \widetilde\alpha_n\frac{D-2n}{D-2} \,,
\end{equation}
in terms of which
\begin{align}
    G_{2}(\varphi, X)&=\varphi^{D-2}\Big((D-1)h(\psi)-2\psi  h'(\psi)\Big) \,,
    \label{eq:G2-def} \\
    G_{3}(\varphi, X)&=2\varphi^{D-3}h'(\psi) \,,
    \label{eq:G3-def} \\
    G_{4}(\varphi, X)&=-\frac{1}{2}\varphi^{D-2}\psi^{\frac{D-2}{2}}\int \dd \psi \, \psi^{-\frac{D}{2}}h'(\psi) \,, \label{eq:G4-def}
\end{align}
where $h'(\psi)\equiv \dd h(\psi)/\dd \psi$, with $\psi$ given by~\eqref{eq:psiXdef} substituted only after the derivatives and the integral have been taken. These functions are not independent and satisfy
\begin{equation}\label{eq:Gidentities}
    G_{4,\varphi} = \frac{1}{2}G_{3} \,,\qquad G_{2,X} = -\frac{1}{2}G_{3,\varphi} \,.
\end{equation}

\subsection{Field equations}
\label{sec:eom}

Varying~\eqref{eq:2daction} with respect to $\gamma^{\mu\nu}$ and $\varphi$ gives the two-dimensional and angular blocks of $\mathcal{E}_{ab}$,
\begin{align}
    \label{eq:eomtr}
    \mathcal{E}_{\mu\nu} &= \frac{16\pi \GN}{\Omega_{D-2} \varphi^{D-2}\sqrt{|\gamma|}} \frac{\updelta I_{\rm 2d}}{\updelta \gamma^{\mu\nu}} \,,\\
    \label{eq:eomij}
    \mathcal{E}_{ij} &= -\frac{g_{ij} 16\pi \GN}{2(D-2)\Omega_{D-2} \varphi^{D-3}\sqrt{|\gamma|}} \frac{\updelta I_{\rm 2d}}{\updelta \varphi} \,,
\end{align}
where $i,j$ are the angular components (so that $g_{ij}$ denotes the components of the metric of the $(D-2)$-dimensional sphere of radius $\varphi$) and $\mu,\nu$ the two-dimensional components. The explicit result from the variation of~\eqref{eq:2daction} is
\begin{align}
    \mathcal{E}_{\mu\nu} & = \frac{D-2}{\varphi^{D-2}} \Big( G_3 g_{\mu[\nu} \nabla_{\rho]} \nabla^{\rho} \varphi - \frac12 G_2 g_{\mu\nu} \Big) \,, \label{eq:einstein-horn}\\
    \mathcal{E}_{ij} & = \frac{g_{ij}}{2\varphi^{D-3}} \Big( \Box \varphi\, G_{3,\varphi} - G_{2,\varphi} - \frac12 G_3R \notag \\
    & \phantom{{}=\frac{g_{ij}}{2\varphi^{D-3}} } -2\nabla_{[\mu} \nabla^\nu \varphi \nabla_{\nu]} \nabla^{\mu} \varphi \, G_{3,X} \Big)\, ,  \label{eq:einstein-angular}
\end{align}
which are related by the generalized Bianchi identity
\begin{equation}
    \nabla^\mu\big(\varphi^{D-2}\mathcal{E}_{\mu\nu}\big) = \varphi^{D-3}\,\partial_\nu\varphi\, g^{ij}\mathcal{E}_{ij} \,,
    \label{eq:bianchirel}
\end{equation}
so that the angular equations follow from the others together with local conservation of $T_{ab}$.

Fixing the gauge
\begin{equation}\label{eq:gaugeNf}
    \dd s_{\gamma}^2 \!=\! -N(t,r)^2 f(t,r)\,\dd t^2 + \frac{\dd r^2}{f(t,r)} \,, \quad \varphi = r \,,
\end{equation}
such that $X=f$ and $\psi=(1-f)/r^2$, the dynamics of arbitrary spherically symmetric matter reduces to
\begin{align}
    \partial_r \big(r^{D-1}h(\psi)\big) &=-\frac{16\pi \GN}{D-2}\,r^{D-2}\,T^t{}_t \,, \label{eq:dyn1} \\
    \frac{\partial_t f}{r f} &=-\frac{16\pi \GN}{(D-2)h'(\psi)}\,T_{tr} \,, \label{eq:dyn2} \\
    \frac{\partial_r N}{r N} &= \frac{8\pi \GN}{(D-2)fh'(\psi)}\big(T^r{}_r-T^t{}_t\big) \,, \label{eq:dyn3}
\end{align}
together with $\nabla^aT_{ab}=0$.

\subsubsection*{Vacuum solutions}

Setting $T_{ab}=0$ in~\eqref{eq:dyn1}--\eqref{eq:dyn3} forces $\partial_t f=0$, $\partial_rN=0$ and $\partial_r\big(r^{D-1}h(\psi)\big)=0$. Since $N=N(t)$, it can be absorbed into a redefinition of the time coordinate, so the most general spherically symmetric solution is static and fixed by a single algebraic equation,
\begin{equation} \label{eq:eom}
    h(\psi) = \frac{2\mathsf{M}}{r^{D-1}} \,,\qquad
    \mathsf{M} \equiv \frac{8\pi\GN M}{(D-2)\Omega_{D-2}} \,,
\end{equation}
with $M$ the ADM mass. This is the Birkhoff theorem for these theories~\cite{Oliva:2010eb,Oliva:2011xu,Cisterna:2017umf}.

\section{Warm-up: thin-shell limit of Vaidya}
\label{sec:vaidya-warmup}

 Here we shall consider a Vaidya solution of Lovelock or QT gravity in which a black hole absorbs a pulse of null matter over a finite duration of advanced time. We shall examine the behavior of this solution when the pulse of radiation becomes distributional. This provides a pedagogical illustration of the discrepancy between the exact thin-shell limit and the distributional prescription adopted in~\cite{DiFilippo:2026jpv}. In particular, this will allow us to motivate  the Vol'pert chain rule for functions of bounded variation which is the main technical ingredient necessary in the junction conditions. 

Let us consider the ingoing Vaidya-type metric,
\be 
    \dd s^2=
    -f_\ve(v,r)\,\dd v^2
    +2\,\dd v\,\dd r
    +r^2\dd\Omega_{D-2}^2 \,,
    \label{eq:vaidya-ansatz}
\ee
and define
\be 
    \psi_\ve(v,r)
    =
    \frac{1-f_\ve(v,r)}{r^2} \,.
\ee
On a smooth, invertible branch of the characteristic function the metric function is determined by
\be     \label{eq:vaidya-eom}
h(\psi_\ve) = \frac{2 \mathsf{M}_\ve(v)}{r^{D-1}} \, .
\ee
We will take
\be 
\mathsf{M}_\ve(v) = \mathsf{M}_- + \Delta \mathsf{M} H_\ve(v) \,, \quad \Delta \mathsf{M} \equiv \mathsf{M}_+ - \mathsf{M}_- \, ,
\ee
with nascent step and delta functions
\be 
 H_\ve(v)= \int_{-\infty}^v \dd w \,\delta_\ve(w) \, , \quad \delta_\ve = \frac{1}{\ve} \rho \left(\frac{v}{\ve}\right) \,,
\ee
where $\rho$ is a non-negative, smooth function of compact support normalized according to
\be 
\int_{-\infty}^{\infty} \dd x\, \rho(x)  = 1 \, .
\ee
Thus, in the limit $\ve \to 0$, we will have $H_\ve \to \Theta$ and $\delta_\ve \to \delta$ in the sense of distributions.\footnote{In the $(v,r)$ chart $g_{vv}=-f_\ve$ becomes discontinuous as $\ve\to0$, but this is an artifact of the chart. Setting $r=r(v,U)$ with $\partial_v r=\tfrac{1}{2} f$ and $r(0,U)=U$, so that $U$ labels outgoing radial null rays, gives $\dd s^2=2\,\partial_U r\,\dd v\,\dd U+r^2\dd\Omega_{D-2}^2$. Here $r$ is continuous, and so is $\partial_U r=\exp\!\big(\tfrac{1}{2}\int_0^v\dd v'\,\partial_r f\big)$, because $\partial_r f$, although discontinuous, remains bounded. The limiting metric is therefore continuous with bounded first derivatives, as required for distributional curvature to be well-defined. Since  our calculations in this example are performed at $\ve>0$ before taking the limit, none of our conclusions are sensitive to this subtlety, and to avoid obscuring the main points we work in the $(v,r)$ chart throughout.}

For every $\ve > 0$ the geometry and its matter source are smooth. To read off the stress tensor, note the mixed $rv$ field equation reads
\begin{align}
    \mathcal{E}^{r}{}_{v}
    &=
    \frac{D-2}{2r^{D-2}}
    \partial_v\!\left(r^{D-1}h(\psi_\ve)\right)
    \notag\\
    &=
    \frac{D-2}{r^{D-2}}\partial_v{\M}_\ve(v)
    =
    8\pi\GN T^{r}{}_{v}\,.
    \label{eq:vaidya-flux-equation}
\end{align}
The source is therefore an ingoing null fluid with $T^r{}_v = T_{vv}$ and
\be \label{eq:distributional_EMT}
T_{vv}^{(\ve)} = \frac{D-2}{8 \pi \GN} \frac{\Delta \mathsf{M}}{r^{D-2}} \,\delta_\ve (v) \, .
\ee
The flux of matter through a sphere of area $A_{D-2} = \Omega_{D-2} r^{D-2}$ is
\be \label{eq:mass_flux}
m_{\rm shell} = \int_{-\infty}^{+\infty} \!\! \!\!\! \!\! \dd v\, A_{D-2} T^r{}_v = \frac{(D-2) \Omega_{D-2}}{8 \pi \GN} \Delta \mathsf{M} \,.
\ee
The pulse of matter therefore carries a fixed energy which is independent of both its width, characterized by $\ve$, and the radius at which it is measured. The thin-shell limit must retain this finite, radius-independent shell mass. We now examine this limit more carefully, by studying the distributional form of the $\mathcal{E}^{r}{}_{v}$.

At fixed $r$, let $\psi_\pm(r)$ denote the static solutions on either side of the pulse, fixed on the chosen branch by $h(\psi_\pm)=2\M_\pm/r^{D-1}$. We define the jumps 
\begin{align} 
    &[\psi]\equiv\psi_+-\psi_- \,,
    \\
    &[h]\equiv h(\psi_+)-h(\psi_-)=\frac{2\Delta\M}{r^{D-1}} \,.
    \label{eq:vaidya-jumps}
\end{align}
Because $\rho$ has compact support, $\psi_\ve$ coincides with $\psi_-$ before the pulse and with $\psi_+$ after it for every $\ve>0$. Therefore, these jumps are independent of the detailed structure of the transition layer. We can then re-write~\eqref{eq:vaidya-eom} as 
\begin{equation}
    h(\psi_\ve)
    =
    h(\psi_-)+[h]\,H_\ve(v) \,.
    \label{eq:vaidya-h-profile}
\end{equation}
As $\ve\to0$ the transition layer collapses onto the null hypersurface $\Sigma=\{v=0\}$, across which $\psi$ jumps by $[\psi]$ and $h(\psi_\ve)\to h(\psi_-)+[h]\,\Theta(v)$.\footnote{In the later sections, we will be starting from a singular hypersurface $\Sigma$ from the outset. We will then make clear $[\psi] = \psi_+|_{\Sigma} - \psi_-|_{\Sigma}$, \ie the difference is taken between the two sides of the singular surface.}  Since differentiation commutes with the distributional limit, we  immediately have
\be \label{eq:vaidya-limit}
\partial_v h(\psi_\ve) \longrightarrow [h] \, \delta(v) \, .
\ee
This is the distributional limit of the quantity that occurs in the first line of~\eqref{eq:vaidya-flux-equation}. The limit is unambiguous and leads to consistent field equations for the $\ve \to 0$ limit of the stress tensor~\eqref{eq:distributional_EMT}.

There is, however, a second way of writing exactly the same quantity. For every nonzero $\ve$, the ordinary chain rule gives
\be 
    \partial_vh(\psi_\ve)
    =
    h'(\psi_\ve)\partial_v\psi_\ve \,.
    \label{eq:vaidya-chain}
\ee
This re-writing is important because it is precisely the treatment of the quantity on the right that becomes subtle in the thin shell calculation. We will comment on how the distributional limit of this quantity should be computed, and compare this with the results of~\cite{DiFilippo:2026jpv}.

\subsection{The Vol'pert chain rule}

Comparing~\eqref{eq:vaidya-limit} and~\eqref{eq:vaidya-chain} tells us what the $\ve \to 0$ limit of~\eqref{eq:vaidya-chain} must be:
\be  h'(\psi_\ve)\partial_v\psi_\ve
    \longrightarrow
    [h]\,\delta(v) \,.
    \label{eq:vaidya-product-limit}
\ee
The subtlety enters when trying to understand the product $h'(\psi)\partial_v\psi$ as a distribution, which is not defined \textit{a priori}. To this end, note that the distributional limit of  $\psi_\ve$ is
\be
    \psi
    =
    \psi_-+[\psi]\,\Theta(v) \,,
\ee
and hence
\be
    \partial_v\psi
    =
    [\psi]\,\delta(v) \,.
    \label{eq:vaidya-psi-limit}
\ee
Thus, if the chain rule is to retain the form
$h'(\psi)\partial_v\psi=\partial_vh(\psi)$ in the distributional limit we must have the identity 
\be
    h'(\psi)\partial_v\psi
    \equiv
    \frac{[h]}{[\psi]}\,\partial_v\psi 
    \label{eq:vaidya-distributional-chain}
\ee
at the location of the thin pulse. The coefficient $[h]/[\psi]$ is the secant slope of $h$ between the two sides of the pulse.
By the fundamental theorem of calculus, it can also be written as
\be 
    \overline{h'(\psi)}
    \equiv
    \int_0^1\dd s\,
    h'\!\left(\psi_-+s\,[\psi]\right) = \frac{[h]}{[\psi]} \,,
    \label{eq:vaidya-volpert-average}
\ee
where the ratio is understood by continuity when $[\psi]=0$. 
The
distributional product is then defined by
\be
    h'(\psi)\partial_v\psi
    \equiv
    \overline{h'(\psi)}\,\partial_v\psi \,.
    \label{eq:vaidya-volpert-chain}
\ee
The quantity $\overline{h'(\psi)}$ is the \textit{Vol'pert average},
and~\eqref{eq:vaidya-volpert-chain} is the scalar
\textit{Vol'pert chain rule}~\cite{Volpert_1967}. We shall have more to say about this below. 

Crucially, note that~\eqref{eq:vaidya-volpert-chain} does not imply
\be
    h'(\psi_\ve)
    \longrightarrow
    \overline{h'(\psi)} \,.
\ee
Away from the thin shell, $h'(\psi)$ takes its ordinary form. What the Vol'pert average specifies is the factor that multiplies $\partial_v \psi$ at the jump, not the limit of
$h'(\psi_\ve)$ by itself. A second crucial point is that the uniqueness of the Vol'pert average follows from the chain rule structure of the product. For every regularization $h'(\psi_\ve) \partial_v \psi_\ve = \partial_v h(\psi_\ve)$ holds identically. This should be contrasted with a product of independently regularized objects, whose limit could depend on the choice of regularization.

\subsection{The trapezoidal approximation}

We can now compare this result with
Ref.~\cite{DiFilippo:2026jpv}. In that work the discontinuous
quantity $G_3 \sim h'(\psi)$ is first replaced by its two endpoint values,
\be 
    h'(\psi_\varepsilon)
    \longrightarrow
    h'(\psi_+)\Theta(v) +  h'(\psi_-)\Theta(-v)\,,
    \label{eq:vaidya-endpoint-split}
\ee
as in their Eq.~(57), and the resulting products are then evaluated
using prescriptions such as $\Theta\delta=\frac{1}{2}\delta$, their Eq.~(63). For the distributional limit of~\eqref{eq:vaidya-chain} above, this approach results in
\begin{equation}
    h'(\psi)\partial_v\psi
    \stackrel{\rm trap}{\approx}
    \frac{h'(\psi_+)+h'(\psi_-)}{2}\,
    [\psi]\,\delta(v) \,.
    \label{eq:vaidya-endpoint-product}
\end{equation}
This prescription differs from the exact thin-shell limit in~\eqref{eq:vaidya-product-limit}.  While~\eqref{eq:vaidya-endpoint-split} is correct away from the shell, it does not determine the limit of the product at the shell. That is because, as $\ve \to 0$, the $\partial_v \psi_\ve$ term becomes increasingly concentrated in a narrow spike and probes $h'(\psi_\ve)$ on a very fine layer. In more mathematical terms, the problem is that weak limits cannot in
general be multiplied, and therefore
\be 
    \lim_{\ve\to0}
    \bigl[h'(\psi_\ve)
    \partial_v\psi_\ve\bigr]
    \neq
    \bigl[\lim_{\ve\to0}h'(\psi_\varepsilon)\bigr]
    \bigl[\lim_{\ve\to0}
    \partial_v\psi_\ve\bigr] \,.
\end{equation}

The discrepancy between the two results has a simple interpretation. Consider the integration of the nascent quantity
\be 
\int_{-\infty}^{+\infty} \dd v \, h'(\psi_\ve) \partial_v \psi_\ve = \int_{\psi_-}^{\psi_+} \dd z \, h'(z) = [h] \, .
\ee 
The integral has the same form irrespective of the choice of nascent $\delta_\ve$ because of its total derivative nature.  The Vol'pert chain rule ensures that this integral has the same value in the distributional limit. On the other hand, the prescription of Ref.~\cite{DiFilippo:2026jpv} amounts to replacing the exact integral by the trapezoidal approximation to it,
\be 
\int_{\psi_-}^{\psi_+} \dd z \, h'(z)
    \stackrel{\rm trap}{\approx}
    \frac{h'(\psi_+)+h'(\psi_-)}{2}\,[\psi] \,.
\ee  
This explains why Ref.~\cite{DiFilippo:2026jpv} finds that their prescription allows for thin shells in GR and Einstein--Gauss--Bonnet gravity. In those cases $h(x)$ is at most quadratic, and the trapezoidal approximation is then exact.

\subsection{Limitations of the trapezoidal approximation}

 Beyond quadratic order, the trapezoidal approximation does not reproduce the exact jump integral. To quantify the discrepancy, note that inserting~\eqref{eq:vaidya-endpoint-product} into~\eqref{eq:vaidya-flux-equation} and comparing with the thin-shell limit of~\eqref{eq:distributional_EMT} amounts to replacing $\Delta\M$ by
\be 
    \Delta\M_{\rm trap}(r)
    \equiv
    \frac{r^{D-1}}{2}\,
    \frac{h'(\psi_+)+h'(\psi_-)}{2}\,[\psi] \, .
    \label{eq:trap-mass}
\ee
The usual error estimate for the trapezoidal rule  gives
\be
    \Delta\M_{\rm trap}-\Delta\M
    =
    \frac{r^{D-1}}{24}\,[\psi]^3\,h'''(\psi_\star) \,,
    \label{eq:trap-error}
\ee
for some $\psi_\star$ between $\psi_-$ and $\psi_+$. There are then two ways in which the trapezoidal approximation fails. First, $\Delta\M_{\rm trap}$ will generically be a function of $r$. This is incompatible with the radius-independent flux~\eqref{eq:mass_flux}, and can produce an apparent obstruction to thin-shell solutions. Second, in some special cases, $\Delta\M_{\rm trap}$ may be independent of $r$ but still differ from $\Delta\M$. In such a case, the trapezoidal prescription appears to allow thin shells to exist, but the assigned mass disagrees with the exact result.

To see an example of the first kind, consider a cubic theory with $h(\psi)=\psi+\alpha_3\psi^3$. Then, applying~\eqref{eq:trap-mass} one finds 
\be 
\Delta\M_{\rm trap}= \Delta\M + \frac{\alpha_3}{4}\,r^{D-1}\,[\psi]^3\, .
\ee
In this case, the trapezoidal approximation leads to a shell mass that depends on $r$. This same discrepancy occurs in the regular black hole models considered in~\cite{DiFilippo:2026jpv}. For an example of the second kind, consider a pure Lovelock theory with $h(\psi) = \alpha \psi^K$ and $K > 2$. In this case,~\eqref{eq:trap-mass} gives (cf.~Eq.~(97) of~\cite{DiFilippo:2026jpv}) 
\be
    \Delta\M_{\rm trap}
    =
    \frac{K}{2}
    \left(\M_+^{\frac{K-1}{K}}+\M_-^{\frac{K-1}{K}}\right)
    \left(\M_+^{\frac{1}{K}}-\M_-^{\frac{1}{K}}\right) \, .
\ee
The shell mass is independent of radius, but for any $K > 2$ it disagrees with the correct result of $\Delta\M$.

In the sections below we will carefully derive the junction conditions for thin shells for the null and non-null cases. Just as in this simple example, we will see that the Vol'pert chain rule leads to consistent junction conditions in all cases. Moreover, in the non-null case, we shall see that the result agrees with the action-based formalism.

\section{Junction conditions for non-null thin shells} \label{sec:nonnull}
Let a non-null spherical domain wall $\Sigma$ split $\mathcal{M}$ in two regions $\mathcal{M}^\pm$. Let $\{e_A^a\}_{A=1}^{D-1}$ be $D-1$ vectors tangent to $\Sigma$ and $n_a$ its normal.

We allow matter both in the bulk of each region and localized on the wall, and split the matter action as $I_{\rm matter} = I_{\rm matter}^\pm + I_{\rm matter}^{\rm boundary}$, so that
\begin{align} \label{eq:matter-bulk}
    T^\pm_{ab} &= -\frac{2}{\sqrt{|g|}} \frac{\updelta I^\pm_{\rm matter}}{\updelta g_\pm^{ab}} \,, \\
    \label{eq:matter-bdy}
    S_{AB} &= -\frac{2}{\sqrt{|h|}} \frac{\updelta I^{\rm boundary}_{\rm matter}}{\updelta h^{AB}} \,,
\end{align}
where capital Latin indices refer to directions tangent to $\Sigma$.

Take a unit normal $n_a$ with $n_an^a=\epsilon=\pm1$, oriented from $\mathcal{M}^-$ to $\mathcal{M}^+$. For any quantity $Q$ defined on both sides of the wall, we write its jump as
\begin{equation}
    [Q] \equiv Q^+|_\Sigma-Q^-|_\Sigma \, .
\end{equation}

We then parametrize $h_{AB} \dd y^A \dd y^B = h_{\tau\tau} \dd \tau^2 + \varphi(\tau)^2 \dd \Omega^2_{D-2}$. Here $\varphi(\tau)$ is the value of $\varphi$ on $\Sigma$, \ie the areal radius of the shell; to avoid extra notation, we use the same symbol for the field and for its value on the wall. Seen from the point of view of $\gamma_{\mu\nu}$ the wall is a curve $\Sigma'$ with induced metric $h_{\tau\tau}\dd\tau^2$. Here $\Sigma = \Sigma'\times \mathbb{S}^{D-2}$.

We also write
\begin{equation}\label{eq:phinK}
    \varphi^\pm_n \equiv n^\mu\partial_\mu\varphi^\pm \,,\qquad
    K^\pm=\nabla_\mu n_\pm^\mu \,,
\end{equation}
for the normal derivative of the dilaton $\varphi$ at $\Sigma$, and the extrinsic curvature $K$ of the curve $\Sigma'$. The latter is not the full trace over $\Sigma$: including the angular contributions, it is
\begin{equation} \label{eq:KDvsK2}
    K^{(D-1)} = K + (D-2)\frac{\varphi_n}{\varphi}\, .
\end{equation}
We also denote by $\Box^h \varphi$ the Laplacian on $\Sigma'$, and define
\begin{equation} \label{eq:defF}
    F_{w}=\int_{0}^{\varphi_n}  \dd z \, G_{w}\left(\varphi, Y+\epsilon z^2\right) \,, \quad Y=h^{\tau\tau} \dot\varphi^2\,,
\end{equation}
where $w=3,4$, $\dot \varphi=\dd\varphi/\dd\tau$, and $h^{\tau\tau}=1/h_{\tau\tau}$.

Moreover, the shell is described in both approaches with a surface stress tensor of perfect fluid on $\Sigma$
\begin{equation} \label{eq:sAB}
    S_{AB} = (\sigma + \epsilon \, p)  u_{A} u_{B} + p \, h_{AB} \,,
\end{equation}
with $u_A$ a unit tangent vector on $\Sigma$. Its components read
\begin{equation}
    S_{\tau\tau} = \sigma \,, \quad S_{ij} = p\, g_{ij} \,,
\end{equation}
which for a timelike wall describes a thin shell with $\sigma$ the surface matter density, $p$ the isotropic pressure and $u_A$ the $(D-1)$-velocity of the fluid. We further define the shell area and mass (the latter being a proper mass only in the timelike case), together with the rescaled mass and pressure,
\begin{align} \label{eq:m-A-def}
    A_{D-2} \equiv \Omega_{D-2}\varphi^{D-2} \,, \quad
    m \equiv A_{D-2}\,\sigma \,, \\
    \{\m,\p\} \equiv \frac{8\pi\GN}{(D-2)\Omega_{D-2}}\,\{m,p\} \,.
\end{align}

\subsection{Distributional approach}

Our first path to the second junction conditions treats the junction as a distributional solution of the field equations, following the standard formalism~\cite{Poisson:2009pwt}. Let $\ell$ denote proper distance (or time) along a congruence of geodesics crossing $\Sigma$ orthogonally, with $\ell<0$ in $\mathcal{M}^-$ and $\ell>0$ in $\mathcal{M}^+$, hence
\begin{equation}
    n_a = \epsilon \partial_a \ell \,, \quad n^a n_a = \epsilon \,,
\end{equation}
and write every field as $Q = Q^+\Theta(\ell) + Q^-\Theta(-\ell)$ with $\Theta$ the Heaviside distribution, which satisfies: $\Theta^2(\pm\ell)=\Theta(\pm\ell)$, $\Theta(\ell)\Theta(-\ell)=0$ and $\dd\Theta(\pm\ell)/\dd\ell = \pm \delta(\ell)$. These identities hold between almost-everywhere-defined functions and we use them only away from $\delta$'s. The product $\Theta(\ell)\delta(\ell)$ is not fixed by the usual rules, and giving it a value is the central issue of this section.

Regarding the matter content, the distributional stress-energy tensor is written as
\begin{equation} \label{eq:tmunudistrib}
    T_{ab} = T_{ab}^+ \Theta(\ell) + T_{ab}^- \Theta(-\ell)  + S_{ab} \,\delta(\ell) \,,
\end{equation}
where $S_{ab}$ is obtained from $S^{ab}=S^{AB} e_A^ae_B^b$ with $S_{AB}$ given in~\eqref{eq:sAB}.

\subsubsection*{First junction condition}

Consider the distributional metric 
\begin{equation}
    g_{ab} = g_{ab}^+ \Theta(\ell) + g_{ab}^- \Theta(-\ell) \,.
\end{equation}
Using $\partial_c\Theta(\pm\ell)=\pm\epsilon\, n_c\,\delta(\ell)$, we have
\begin{align}
    \partial_c g_{ab} & = \partial_c g_{ab}^+ \Theta(\ell) + \partial_c g_{ab}^- \Theta(-\ell) \notag \\
    &\phantom{={}}+ \epsilon\, n_c \,[g_{ab}]\, \delta(\ell) \,.
\end{align}
Were the last term allowed to survive, the Christoffel symbols would contain a $\delta$. Then, the curvature tensor would contain both $\delta'$ and $\delta~\times~\delta$.  The connection would then not be defined as a distribution at all. Imposing $[g_{ab}]=0$ removes such a term, and its coordinate-invariant content is
\begin{equation}
    [h_{AB}] = 0 \qquad\Longrightarrow\qquad [h_{\tau\tau}]=[\varphi]=0 \,.
\end{equation}
Since the metric is continuous, the tangential derivatives are continuous as well, so any discontinuity in $g_{ab,c}$ points along the normal~\cite{Poisson:2009pwt}. Regarding the angular block of the metric, one has
\begin{equation}\label{eq:only-phin-jumps}
    [\partial_a\varphi] = \epsilon\,[\varphi_n]\, n_a \,.
\end{equation}
Hence, $\varphi_n$ is the only scalar built from $\partial\varphi$ that jumps across $\Sigma$. The two-dimensional metric  carries its own discontinuity, $[K]\neq0$, and a relation between the two jumps will emerge below.

\subsubsection*{Structure of the singular terms}

The objects entering the field equations fall into three classes:
\begin{center}
    \begin{tabular}{ll}
        \hline
        continuous & $\quad g$, $\varphi$ \\
        bounded & $\quad \partial g$, $\partial\varphi$ \\
        bounded part $\times\delta$ & $\quad\partial\partial g$, $\partial\partial\varphi$ \\
        \hline
    \end{tabular}
\end{center}
where the bounded part is allowed to have a jump.

No $\delta'$ or higher-order distribution appears: the field equations are second order, the metric is continuous, and only its first derivatives jump, so the highest singular degree they can reach is $\delta$. The singular content is automatically of order zero, and the surface stress tensor is a Radon measure, integrating to finite values across the wall. Products of two $\delta$'s are likewise absent --- such a term is ill-defined as a distribution and would require Colombeau algebras to make sense of it~\cite{Steinbauer:2006qi,Chu:2021uec,Silva:2026jiv}. The only term in~\eqref{eq:einstein-horn} and~\eqref{eq:einstein-angular} quadratic in second derivatives is the $G_{3,X}$ one in~\eqref{eq:einstein-angular}. Since the singular part of the Hessian is
\begin{equation} \label{eq:hess-dilaton}
    \nabla_\mu\nabla_\nu\varphi|_{\rm sing.} = n_\mu n_\nu \, [\varphi_n] \, \delta(\ell) \,,
\end{equation}
the antisymmetrization kills the piece where both factors are singular; only the cross term survives, and we evaluate it below.

A bounded quantity with a jump can, however, multiply the $\delta$. In GR this never occurs: the field equations are linear in the second derivatives, and the coefficients multiplying them are continuous. Here they are not. In~\eqref{eq:einstein-horn} the $\delta$ produced by $\nabla\nabla\varphi$ comes multiplied by $G_3(\varphi,X)$, which jumps across $\Sigma$ because $X$ depends on $\varphi_n$. The value of $G_3$ on the wall is therefore undetermined, and the $\delta$ picks out exactly the point where it is. 

Since all discontinuous quantities here are functions of the same coordinate $\ell$, every $\delta$ appearing in the equations is $\delta(\ell)$. This alone does not give the products a definite value; what does is the structure of the combinations in which they appear.

\subsubsection*{Fixing the singular products: the Vol'pert chain rule}

The difficulty is not the multiplication itself: a function integrable with respect to a measure can be multiplied by it. The problem resides, however, in Vol'pert's words~\cite{Volpert_1967}, in ``assigning values to functions on sets on which the singular parts of measures are concentrated.'' In~\eqref{eq:einstein-horn} the $\delta$ is supported on $\Sigma$, and that is precisely where $G_3$ has no definite value.

The pointwise substitution $\Theta^n=\Theta$ for every $n$ does not work. Using the ordinary composition in the chain rule would give $\partial_\ell\Theta = n\,\Theta^{n-1}\,\partial_\ell\Theta = n\,\Theta \, \partial_\ell\Theta= \frac{n}{2} \partial_\ell \Theta$, which cannot hold for all $n$. What does work, as Vol'pert showed, is to replace it by the \emph{averaged composition}.\footnote{Proofs can be found in the original work~\cite{Volpert_1967} and a compact presentation is given in~\cite{Ranocha_2018}.}

\textbf{Scalar Vol'pert chain rule.}
Let $u$ be a function of bounded variation --- for our purposes, a piecewise-smooth function with finite jumps --- so that it admits finite one-sided limits $u^\pm$ on either side of the wall. The Vol'pert average of a continuous function $f$ along the segment joining them is
\begin{equation}\label{eq:volpert-average}
    \overline{f(u)} \equiv \int_0^1 \dd s\, f\big(u^- + s\,[u]\big) \,.
\end{equation}
The chain rule then survives in the distributional sense,
\begin{equation}\label{eq:volpert-dif}
    \partial_\ell f(u) = \overline{f'(u)}\,\partial_\ell u \,.
\end{equation}
The example above is now consistent: $\overline{\Theta^{n-1}} = \int_0^1 \dd s\, s^{n-1} = 1/n$ cancels the factor $n$, and one recovers $\partial_\ell\Theta$ for every power. Equivalently,
\begin{equation} \label{eq:thetadeltarule}
    \overline{\Theta^{n-1}}\,\delta(\ell) = \delta(\ell)/n \,.
\end{equation}
This relation also appears in Ref.~\cite{DiFilippo:2026jpv}. Appendix~\ref{app:resummation} shows how its term-by-term application reproduces the exact junction conditions.

Away from the jump the two limits coincide, the average is simply $f'(u)$, and~\eqref{eq:volpert-dif} is the ordinary chain rule. Its content lies in the singular part, where it states that
\begin{equation}\label{eq:volpert-jump}
    [f(u)] = \int_0^1 \dd s\, \frac{\dd}{\dd s}f\big(u(s)\big) = \overline{f'(u)}\,[u] \,,
\end{equation}
as follows by parameterizing the segment as $u(s)=u^-+s\,[u]$, with $\dd u/\dd s = [u]$. The coefficient multiplying the $\delta$ is thus the mean of $f'$ across the wall.

In our case, in particular, the jumping variable is the normal derivative of the dilaton, $u = \varphi_n^- + [\varphi_n]\,\Theta(\ell)$, and the $\delta$ in the field equations is not an independent object: it is $\delta(\ell)=\partial_\ell u/[\varphi_n]$. The singular terms therefore have the form $f'(u)\,\partial_\ell u$, and~\eqref{eq:volpert-dif} applies directly: when $\varphi_n$ is the only quantity that jumps, the discontinuous coefficient multiplying the $\delta$ is replaced by its Vol'pert average. Note that since $f'$ always has a primitive, the result is the jump of a potential, $[f(u)]\,\delta(\ell)$, and does not depend on the profile of $u$ inside the layer. The vectorial form of the rule, which applies when more than one quantity jumps and is needed for the angular equation, is given below.

\textbf{Vectorial Vol'pert chain rule.} 
Let $u=(u_1,\dots,u_m)$ be a vector of functions of bounded variation with one-sided limits $u^\pm \in \mathbb{R}^m$, and let the averaged composition of a continuous $g:\mathbb{R}^m\to\mathbb{R}$ be taken along the straight segment $u(s)=u^-+s\,[u]$, $s\in[0,1]$, joining the two limiting values, so that all components vary simultaneously and linearly,
\begin{equation} \label{eq:avg-vector}
    \overline{g(u)} \equiv \int_0^1 \dd s \, g \big( u^- + s \,[u]\big) \,.
\end{equation}

Then, for $f \in C^1(\mathbb{R}^m; \mathbb{R})$, the Vol'pert chain rule is
\begin{equation}\label{eq:chain-vector}
    \partial_\ell f(u) = \sum_{k=1}^{m} \overline{\partial_{u_k} f(u)} \, \partial_\ell u_k \,,
\end{equation}
which reduces to~\eqref{eq:volpert-dif} for $m=1$. At the jump point it states
\begin{equation}\label{eq:jump-vector}
    [f(u)]=\sum_{k=1}^{m}\overline{\partial_{u_k}f(u)}\,[u_k]\,.
\end{equation}

As in the scalar case, the right-hand side is the differential of a single \textit{potential} function $f$, but with several variables this is no longer straightforward. When $\sum_i c_i(u) \partial_\ell u_i$ is not an exact differential, its jump is profile dependent and the linear parametrization used in the Vol'pert average becomes one of the possible choices. For instance, $\int v\,\dd u$ between $(u,v)=(0,0)$ and $(1,1)$ depends on the path, whereas $\int(v\,\dd u+u\,\dd v)=[uv]$ does not. Definitions for such non-conservative products have been investigated in~\cite{DalMaso1995DefinitionAW}. Since the singular combinations appearing in the field equations turn out to be exact differentials, we will use the Vol'pert chain rule as a convenient way to compute them.

A useful particular case follows immediately. Setting $m=2$ and taking $f(u_1,u_2)=f_1(u_1)\,u_2$, so that $\partial_{u_1}f = f_1'(u_1)\,u_2$ and $\partial_{u_2}f = f_1(u_1)$, the jump relation~\eqref{eq:jump-vector} gives
\begin{equation}\label{eq:product-vector}
    \big[f_1(u_1)\,u_2\big] = \overline{f_1'(u_1)\,u_2}\,[u_1] + \overline{f_1(u_1)}\,[u_2] \,.
\end{equation}

\subsubsection*{Second junction condition: the two-dimensional components}

In this approach, the second junction conditions come from matching the singular parts of the two sides of the field equations. To begin with, the singular part of~\eqref{eq:einstein-horn} comes only from $\nabla\nabla\varphi$ and reads
\begin{equation}\label{eq:Emunusing}
    \mathcal{E}_{\mu\nu}|_{\rm sing.} = \frac{(D-2)\epsilon}{2\varphi^{D-2}} h_{\mu\nu} \overline{G_3} \,[\varphi_n]\, \delta(\ell) \,,
\end{equation}
where the bar denotes the Vol'pert average over the only non-continuous quantity across $\Sigma$ and we identify $[\varphi_n]\, \delta(\ell)$ as $(\partial_\ell \varphi_n)|_{\rm sing.}$. 

On the wall the singular stress tensor is $T_{ab}|_{\rm sing.} = S_{ab} \,\delta(\ell)$, so projecting the field equations along $\tau\tau$ with $S_{\tau\tau}=\sigma$,
\begin{equation} \label{eq:Emunusing2}
    \frac{(D-2)\epsilon}{2\varphi^{D-2}} h_{\tau\tau} \overline{G_3} \,[\varphi_n]\, \delta(\ell)  = 8 \pi \GN \sigma\, \delta(\ell) \,.
\end{equation}
The gauge choice $h_{\tau\tau}=-\epsilon$ cancels the two factors of $\epsilon$, and using~\eqref{eq:m-A-def}, we are left with
\begin{equation}\label{eq:eom-thetadelta}
    - \overline{G_3} \, [\varphi_n] \,\delta(\ell)  = 2 \m\, \delta(\ell) \,.
\end{equation}

The left-hand side is the product discussed above: $G_3(\varphi,X)$ is a function of $\varphi_n$ through $X=-\epsilon\dot{\varphi}^2 + \epsilon\varphi_n^2$, and it multiplies a $\delta$ located where $\varphi_n$ jumps. We evaluate the product in two ways: (i) by the Vol'pert chain rule and (ii) by identifying the potential. For the sake of completeness, Appendix~\ref{app:resummation} recovers the same result by resumming the series with $\Theta^{n-1}\delta=\delta/n$ as a prescription.

\textbf{(i) By the Vol'pert chain rule.}
The average~\eqref{eq:volpert-average} of $G_3$ along the jump of $\varphi_n$ reads
\begin{align}
    \overline{G_3} & = \int_0^1 \dd s \, G_3\big(\varphi,-\epsilon\dot\varphi^2 +\epsilon (\varphi_n^- + s\,[\varphi_n])^2 \big) \notag \\
    & = \frac{1}{[\varphi_n]}\int_{\varphi_n^-}^{\varphi_n^+}\dd z \, G_3\!\left(\varphi, -\epsilon\dot\varphi^2 + \epsilon z^2\right) \,,
\end{align}
the second equality following from the change of variable $z=\varphi_n^- + s\,[\varphi_n]$. Note that the last integral is $[F_3]$, with $F_3$ defined in~\eqref{eq:defF}, so this is~\eqref{eq:volpert-jump} with $f=F_3$ and
\begin{equation} \label{eq:barG3}
    \overline{G_3} = \frac{[F_3]}{[\varphi_n]} \,.
\end{equation}
Plugging this into the field equation,
\begin{equation} \label{eq:jumpf3}
    -[F_3] = 2 \m \,,
\end{equation}
that has the following integral form
\begin{equation}\label{eq:2jc-volpert-bulk}
    \int_{\varphi_n^+}^{\varphi_n^-} \dd z \, h'\left(\frac{1+\epsilon \dot{\varphi}^2 - \epsilon z^2}{\varphi^2}\right) = \frac{\m}{\varphi^{D-3}} \,,
\end{equation}
where we have reintroduced $\psi$ using~\eqref{eq:psiXdef}. Here $\varphi_n^\pm$ is fixed by matching to the bulk solution on each side. This equation is known as \textit{the shell-equation}.

\textbf{(ii) By identifying the potential.}
We now rewrite the singular part of the two-dimensional field equation~\eqref{eq:eom-thetadelta} as an exact differential (without the Vol'pert average) using $[\varphi_n]\, \delta(\ell) = (\partial_\ell \varphi_n)|_{\rm sing.} $ and $\partial F_3/\partial\varphi_n = G_3$,
\begin{align}
    -G_3 (\partial_\ell \varphi_n)|_{\rm sing.} & = 2\m \, \delta(\ell) \,, \\
    -(\partial_\ell F_3) |_{\rm sing.} & = 2\m\, \delta(\ell) \,,
\end{align}
so $F_3$ is the \textit{potential function} in the sense that the singular part can be obtained by its differentiation. We then denote
\begin{equation} \label{eq:pot-nonull-tautau}
    \Phi_{\tau\tau} = F_3 \,,
\end{equation}
which is defined up to a function continuous across the shell.

For the sake of completeness, integrating the last expression in $\ell$ over an infinitesimal interval $(-\Delta/2,\Delta/2)$ and taking the limit $\Delta\to0$, we find the same result, $-[\Phi_{\tau\tau}] = 2\m$.

\subsubsection*{Second junction condition: the angular components}

Regarding the angular part, the singular contributions to~\eqref{eq:einstein-angular} follow from the Hessian of the dilaton~\eqref{eq:hess-dilaton} and from the curvature, whose singular part is built from the jump of the extrinsic curvature. They are
\begin{align}
    \Box\varphi\big|_{\rm sing.} &= \epsilon \,[\varphi_n]\, \delta(\ell) \,, \label{eq:sing-box} \\
    R\big|_{\rm sing.} &= -2 \epsilon \,[K]\,\delta(\ell) \,, \label{eq:sing-R} \\
    \nabla_{[\mu} \nabla^{\nu} \varphi \nabla_{\nu]} \nabla^{\mu} \varphi |_{\rm sing.} &= -\epsilon h^{\mu\nu}\nabla_\mu \nabla_\nu \varphi|_{\rm reg.} \, [\varphi_n] \,\delta(\ell) \,, \label{eq:sing-boxbox}
\end{align}
where the label ``reg.'' denotes the regular part,
\begin{equation}
    \nabla_\mu\nabla_\nu\varphi|_{\rm reg.} = \Theta(\ell)\,\nabla_\mu\nabla_\nu\varphi^+ + \Theta(-\ell)\,\nabla_\mu\nabla_\nu\varphi^- \,,
\end{equation}
and the last contribution is the cross term announced above, linear in $\delta$. The tangential Hessian splits as
\begin{equation}\label{eq:hess-tangential}
    h^{\mu\nu}\nabla_\mu\nabla_\nu\varphi = \Box^h\varphi + \epsilon K\varphi_n \,,
\end{equation}
where $\Box^h\varphi$ is continuous across the wall, while $K\varphi_n$ jumps. Not every $\delta$ descends from $\varphi_n$: the one in~\eqref{eq:sing-R} comes with $[K]$. 

\textbf{(i) By the Vol'pert chain rule.}
Since the complete singular combination is an exact differential in $(\varphi_n,K)$, as shown in (ii) below, the vectorial Vol'pert chain rule applies, and in particular~\eqref{eq:product-vector} for $G_3$ and $K$. Imposing $\mathcal{E}_{ij}|_{\rm sing.} = 8\pi\GN p g_{ij} \, \delta(\ell)$ then gives
\begin{equation} \label{eq:angular-jumps}
    \overline{G_{3,\varphi}} \,[\varphi_n] + 2\, \Box^{h}\varphi \, \overline{G_{3,X}} \,[\varphi_n] + [G_3 K] = 2\epsilon \p \frac{\dd A_{D-2}}{\dd \varphi} \,.
\end{equation}

In the GR limit $G_{3,X}=0$ and $G_3=2\varphi^{D-3}$ is continuous, so~\eqref{eq:angular-jumps} collapses to
\begin{equation}\label{eq:angular-GR}
    [K] + (D-3)\frac{[\varphi_n]}{\varphi} =  8\pi\GN \epsilon p\,,
\end{equation}
which by~\eqref{eq:KDvsK2} is $[K^{(D-1)}] = [\varphi_n]/\varphi + 8\pi\GN\epsilon p$, the angular Israel condition for a thin shell~\cite{Israel:1966rt,Poisson:2009pwt}.

\textbf{(ii) By identifying the potential.}
Alternatively, the singular part of the angular equation can be written directly as an exact differential in $(\varphi_n,K)$. After using
\begin{align}
    [\varphi_n] \, \delta(\ell) &= (\partial_\ell \varphi_n)|_{\rm sing.} \,, \\
    [K] \, \delta(\ell) &= (\partial_\ell K)|_{\rm sing.} \,, \\
    (\partial_\ell G_3 )|_{\rm sing.} & = 2\epsilon G_{3,X}\varphi_n (\partial_\ell \varphi_n)|_{\rm sing.} \,, \\
    (\partial_\ell F_{3,\varphi})|_{\rm sing.} &= G_{3,\varphi} (\partial_\ell \varphi_n)|_{\rm sing.} \,, \\
    (\partial_\ell F_{3,Y})|_{\rm sing.} &= G_{3,X} (\partial_\ell \varphi_n)|_{\rm sing.} \,,
\end{align}
imposing $\mathcal{E}_{ij}|_{\rm sing.} = 8\pi\GN p\, g_{ij}\,\delta(\ell)$ gives the exact differential
\begin{equation}
    \partial_\ell \big( G_3 K + F_{3,\varphi} + 2 \, \Box^h\varphi \,F_{3,Y} \big)|_{\rm sing.} = 2\epsilon\p \frac{\dd A_{D-2}}{\dd \varphi} \, \delta(\ell) \,.
\end{equation}
Therefore, the potential reads
\begin{equation}\label{eq:pot-nonull-varphi}
    \Phi_{\varphi}(\varphi_n,K) = -\epsilon(G_3 K + F_{3,\varphi} + 2\, \Box^{h} \varphi\, F_{3,Y}) \,.
\end{equation}
Integrating then in $\ell$ over an infinitesimal interval across the shell, only the singular terms contribute, and~\eqref{eq:angular-jumps} is simply
\begin{equation}\label{eq:pot-nonull-varphi-jump}
    -[\Phi_\varphi] = 2\p\frac{\dd A_{D-2}}{\dd\varphi} \,.
\end{equation}
As for the two-dimensional potential $\Phi_{\tau\tau}$, the dilatonic one, $\Phi_\varphi$, is defined up to a continuous function across the wall.

\subsubsection*{Mass conservation}

The energy balance on $\Sigma$ follows from conservation of the stress tensor alone, without using the junction conditions. Taking the distributional $T^{ab}$ defined in~\eqref{eq:tmunudistrib} and using $\partial_a\Theta(\pm\ell) = \pm \epsilon n_a \delta(\ell)$ together with $\partial_a\delta(\ell) = \epsilon n_a \delta'(\ell)$, its divergence is
\begin{align}
    \nabla_a T^{ab} & = \nabla_a T^{ab}_+ \Theta(\ell) + \nabla_a T^{ab}_- \Theta(-\ell) \notag \\
    & \phantom{{}=} + \epsilon n_a \, [T^{ab}]\, \delta(\ell) + \nabla_a S^{ab} \, \delta(\ell) \notag \\
    & \phantom{{}=} + \epsilon n_aS^{ab} \, \delta'(\ell) \,.
\end{align}
The first two terms vanish separately, due to the conservation law in both regions. The higher-order distribution term, $\delta'(\ell)$, also disappears, since $S_{AB}$ is a surface tensor tangent to $\Sigma$ by construction, so $S^{ab}n_a = S^{AB}(e^a_A n_a)e^b_B = 0$ identically, for any surface source, and this condition holds in a
neighborhood of $\Sigma$ for a tangential continuation of $S^{ab}$.

What remains once we demand conservation is
\begin{equation}
    \nabla_a S^{ab}\, \delta(\ell) +  \epsilon n_a \,[T^{ab}]\, \delta(\ell) = 0 \,,
\end{equation}
whose projection along $u_b$ gives
\begin{equation} \label{eq:second-jc-continuity-distrib}
    \frac{\dd m}{\dd \tau} +\epsilon p \frac{\dd{A}_{D-2}}{\dd \tau} - A_{D-2} \,[T_{\tau n}] = 0\,,
\end{equation}
with $T_{\tau n}\equiv T_{ab} u^a n^b$. The second term is the work done by the pressure against the change in area, so that the shell mass is conserved only for a pressureless shell with no flux across it,
\begin{equation}
    p = [T_{\tau n}] = 0 \qquad \Longrightarrow \qquad m = \text{constant} \,.
\end{equation}

\subsubsection*{The singular part of the Lagrangian: the boundary term}

To close this section we perform a consistency check that ties the two routes together. Plugging the singular parts~\eqref{eq:sing-box},~\eqref{eq:sing-R} and~\eqref{eq:sing-boxbox} into the two-dimensional Lagrangian~\eqref{eq:horn} one finds, after using~\eqref{eq:hess-tangential} and applying~\eqref{eq:product-vector} for $G_4$ and $K$,
\begin{align}
    & \mathcal{L}_{\rm 2d}|_{\rm sing.} \notag \\
    & = - \epsilon \Big( \overline{G_{3}}\,[\varphi_n] + 2 \,[G_4K] + 4 \, \Box^h\varphi \overline{G_{4,X}} \,[\varphi_n] \Big) \delta(\ell) \notag \\
    & = - \epsilon \, \partial_\ell \Big( F_3+ 2 G_4 K + 4 \, \Box^h\varphi F_{4,Y} \Big)\Big|_{\rm sing.} \,, \label{eq:horn-sing}
\end{align}
where in the last line we have used $\overline{G_3} \, [\varphi_n] = [F_3]$ and $\overline{G_{4,X}} \, [\varphi_n] = [F_{4,Y}]$. The singular part of the Lagrangian is thus an exact differential.

In the $\tau,\ell$ coordinates the volume element factorizes as $\dd^2 x \sqrt{|\gamma|} = \dd\tau \, \dd\ell \, \sqrt{|h_{\tau\tau}|}$, so the $\ell$ integral is trivial and gives
\begin{align} \notag
    &\int \dd^2 x \sqrt{|\gamma|} \, \mathcal{L}_{\rm 2d}|_{\rm sing.} \\
    & = - \left[\epsilon \int \dd \tau \, \sqrt{|h_{\tau\tau}|} ( F_3+ 2 G_4 K + 4 \,\Box^h\varphi F_{4,Y} ) \right] \,. \label{eq:horn-sing-action}
\end{align}
It can be written as
\begin{equation}
    \int \dd^2 x \sqrt{|\gamma|} \, \mathcal{L}_{\rm 2d}|_{\rm sing.} = - B^+ + B^- \,,
\end{equation}
where the boundary term $B$, whose integrand is the \textit{potential} of the singular part of the Lagrangian, is defined as
\begin{equation} \label{eq:b-term}
    B = \epsilon \int \dd \tau \, \sqrt{|h_{\tau\tau}|} ( F_3+ 2 G_4 K + 4\, \Box^h\varphi F_{4,Y}) \,.
\end{equation}

Taking into account that the outward normal of each region $\mathcal{M}^\pm$ is $\mp n$, so that $-B^+ = B^+_{(-n)}$, we can write
\begin{align}
    &I^+_{\rm reg.} + I^-_{\rm reg.} + I_{\rm sing.} \notag \\
    &= I^+_{\rm reg.} + I^-_{\rm reg.} + \frac{(D-2)\Omega_{D-2}}{16\pi\GN} \big( B_{(-n)}^+ + B_{(+n)}^- \big)\,,
\end{align}
where $I^\pm_{\rm reg.}$ is the bulk action in $\mathcal{M}^\pm$ and $I_{\rm sing.}$ the total singular contribution.

At this point one can abandon the distributional description in favor of a regular problem on each side and consider the variational problem, where the variation of each $B$ cancels the total derivatives produced by the variation of the bulk action.

\subsection{Variational approach using boundary terms}\label{sec:Junction}

The second route to the junction conditions goes through the boundary terms in the variation of the action. It was used in~\cite{Bueno:2024zsx} to study the gravitational collapse of a dust thin shell, obtaining regular black holes as byproducts. Here we extend that treatment to thin shells with pressure and to configurations with matter on either side of the wall, which allows for fluxes of matter across it.

\subsubsection*{First junction condition}

As in the distributional approach, the first junction condition is the continuity of the induced metric,
\begin{equation} \label{eq:firstJC}
    [h_{AB}]=0 \qquad \Longrightarrow \qquad [h_{\tau\tau}]=[\varphi]=0 \,.
\end{equation}

\subsubsection*{Boundary equations of motion}

This requires supplementing the action~\eqref{eq:QTaction} with a generalized York--Gibbons--Hawking boundary term,
\begin{equation}\label{Stotal1}
    I^{\rm total} = I^{\rm QT} + I^{\rm boundary} \,,
\end{equation}
that makes the variational problem well-posed. The total variation of the gravitational action then is
\begin{equation}\label{variationStotal}
\begin{aligned} 
    \updelta I^{\rm total}&=+\frac{1}{16\pi \GN}\int_\mathcal{M} \dd^{D}x\sqrt{|g|} \, \mathcal{E}_{ab}\updelta g^{ab}\\
    & \phantom{{}=}+\frac{1}{16\pi \GN}\int_{\Sigma} \dd^{D-1}y\sqrt{|h|} \, \Pi_{AB}\updelta h^{AB} \,.
\end{aligned}
\end{equation}
Here $\mathcal{E}_{ab}$ are the bulk field equations~\eqref{eq:Eab}, while $\Pi_{AB}$ represents the boundary equations of motion, which are given by
\begin{equation}
    \Pi_{AB}=\frac{16\pi \GN}{\sqrt{|h|}}\frac{\updelta I^{\rm total}}{\updelta h^{AB}}\, .
\end{equation}
Demanding~\eqref{variationStotal} together with~\eqref{eq:matter-bulk} and~\eqref{eq:matter-bdy} to be stationary gives the bulk field equations on either side and, on the wall itself, the second junction condition~\cite{Israel:1966rt, Brown:1992br, Davis:2002gn, Padilla:2012ze},
\begin{equation}\label{eq:2ndjunction}
    -[\Pi_{AB}]=8\pi \GN S_{AB}\, .
\end{equation}
The minus sign is a matter of orientation.\footnote{$\Sigma$ is the common boundary of $\mathcal{M}^+$ and $\mathcal{M}^-$, whose outward normals are $-n$ and $+n$ respectively, and since $\Pi_{AB}$ is odd under $n\to-n$ the two contributions add up to a jump.}

The hard part is finding the boundary terms, since higher-derivative theories do not in general admit a well-defined variational problem. QT gravities and spherical symmetry solve this for us: the reduced action~\eqref{eq:2daction} is a two-dimensional Horndeski theory, whose equations are second order and whose boundary terms are known~\cite{Padilla:2012ze}. We have recomputed them for~\eqref{eq:horn} and find agreement up to a factor of $\epsilon$, irrelevant for timelike walls but not for spacelike ones. Its origin is the surface element entering Gauss's theorem, $\dd\Sigma_a=\epsilon\,n_a\sqrt{|h|}\,\dd^{D-1}y$, which the boundary term inherits when the total derivative produced by the variation is turned into a surface integral. The same factor appears in the York--Gibbons--Hawking term of GR. In agreement with the result obtained via the distributional method --- see~\eqref{eq:b-term} --- the boundary term carries an explicit factor of $\epsilon$.

The total two-dimensional action reads
\begin{align}
    \label{eq:actot2d}
    & I_{\rm 2d}^{\rm total} = I_{\rm 2d} + \frac{(D-2) \Omega_{D-2}}{16 \pi \GN}B\,,\\
    & B\!=\!\epsilon \! \int_{\Sigma'} \!\dd \tau \sqrt{\vert h_{\tau\tau} \vert} \! \left(F_3+2G_4 K +4 \,\Box^h \varphi F_{4,Y}\right) \!, \label{eq:b}
\end{align}
with $F_3$ and $F_4$ defined in~\eqref{eq:defF}.

Writing the variation of~\eqref{eq:actot2d} as
\begin{equation}
    \begin{aligned}
        \updelta I^{\rm total}_{\rm 2d}& = +\int \dd^{2}x \left( \frac{\updelta I_{\rm 2d}}{\updelta \gamma^{\mu\nu}} \updelta \gamma^{\mu\nu} + \frac{\updelta I_{\rm 2d}}{\updelta\varphi} \updelta \varphi\right) \\
        & \phantom{{}=} + \int_{\Sigma'} \dd \tau \sqrt{|h_{\tau\tau}|} \left(J_{\tau\tau} \updelta h^{\tau\tau} + J_{\varphi} \updelta\varphi \right) \,,
    \end{aligned}
\end{equation}
and comparing with~\eqref{variationStotal}, we recover~\eqref{eq:eomtr} and~\eqref{eq:eomij} in the bulk, and fix the boundary tensor on $\Sigma$ to be
\begin{align}
    \label{eq:eombtr}
    \Pi_{\tau\tau} &=\frac{16\pi \GN}{\Omega_{D-2} \varphi^{D-2}}J_{\tau\tau} \,,\\
    \label{eq:eombij}
    \Pi_{ij} &=-\frac{g_{ij} 16\pi \GN}{2(D-2)\Omega_{D-2} \varphi^{D-3}}J_{\varphi} \,.
\end{align}
Extracting $J_{\tau\tau}$ and $J_\varphi$ from~\eqref{eq:actot2d} is a tedious computation; we only present here the final result after setting $h_{\tau\tau}=-\epsilon$, such that $\tau$ is the proper time or length. We find 
\begin{align}
    J_{\tau\tau} & = \frac{(D-2)\Omega_{D-2}}{16\pi\GN} \frac12 F_3 \,, \\
    J_{\varphi} & = \frac{(D-2)\Omega_{D-2}}{16\pi\GN} \frac{\epsilon}{\dot{\varphi}} \left( \frac{\dd F_3}{\dd \tau} - G_3 \nabla_n \nabla_\tau \varphi \right) \,,
\end{align}
where $\nabla_n \nabla_\tau \varphi \equiv n^\mu u^\nu \nabla_\mu \nabla_\nu \varphi = \dot{\varphi}_n -K \dot{\varphi}$. The second term of $J_\varphi$ can be traded for the momentum flux through the wall using the field equations,
\begin{equation}\label{eq:extra-term}
    -G_3 \nabla_n \nabla_\tau \varphi = \frac{2 \varphi^{D-2}}{D-2} \mathcal{E}_{\tau n} = \frac{2 \varphi^{D-2}}{D-2} 8 \pi\GN T_{\tau n} \,.
\end{equation}
Hence, this term vanishes on either side of the shell whenever there is no momentum flux through it, as in vacuum.

Thus
\begin{align}
    \label{eq:eom2btr}
    \Pi_{\tau\tau}&=\frac{(D-2)}{2\varphi^{D-2}}F_{3}\,, \\
    \label{eq:eom2bij}
    \Pi_\varphi \equiv g^{ij} \Pi_{ij} &=-\frac{\epsilon}{\varphi^{D-3}\dot{\varphi}} \Bigg( \frac{\dd}{\dd \tau} \left(\varphi^{D-2} \Pi_{\tau\tau}\right) \notag \\ 
    & \phantom{{}=} + \varphi^{D-2} 8\pi\GN T_{\tau n} \Bigg) \,.
\end{align}
Using~\eqref{eq:G3-def}, the expression for $\Pi_{\tau\tau}$ in~\eqref{eq:eom2btr} gets simplified into
\begin{equation}\label{eq:Pitautau}
    \Pi_{\tau\tau}=\frac{D-2}{\varphi}\int_{0}^{\varphi_n} \dd z \, h'\left(\frac{1+\epsilon \dot\varphi^2-\epsilon z^2}{\varphi^2}\right)\, .
\end{equation}

\subsubsection*{Second junction conditions for a thin shell}

Consider the perfect-fluid shell~\eqref{eq:sAB}. As stated at the beginning of this section, we do not assume that the shell is the only matter content of the spacetime; there may be stress-energy tensors $T_{ab}^\pm$ on either side. The vacuum case reduces to the thin-shell collapse studied in~\cite{Bueno:2024zsx}.

The boundary equations of motion~\eqref{eq:2ndjunction} read, in components,
\begin{align}\label{junctiontau}
    -[\Pi_{\tau\tau}] &= 8\pi \GN \sigma \,,\\
    -[\Pi_{ij}] & = 8 \pi \GN p g_{ij} \,. \label{eq:junctionij}
\end{align}
We notice here that the potentials $\Phi_{\tau\tau}$ and $\Phi_{\varphi}$ obtained in~\eqref{eq:pot-nonull-tautau} and~\eqref{eq:pot-nonull-varphi} are, up to a normalization, the boundary momenta
\begin{align}
    \Pi_{\tau\tau} & = \frac{D-2}{2\varphi^{D-2}} \Phi_{\tau\tau} \,, \\
    \Pi_{\varphi}  & = \frac{D-2}{2\varphi^{D-3}} \Phi_{\varphi}\,.
\end{align}

\subsubsection*{Second junction condition: the two-dimensional components}

Inserting~\eqref{eq:Pitautau} into~\eqref{junctiontau}, the second junction condition takes the form of a shell-equation,
\begin{equation}\label{shellequation1}
    \int_{\varphi_n^{+}}^{\varphi_n^{-}} \dd z \, h'\left(\frac{1 + \epsilon \dot{\varphi}^2 - \epsilon z^2}{\varphi^2}\right) = \frac{\m}{\varphi^{D-3}}  \,.
\end{equation}
This is exactly~\eqref{eq:2jc-volpert-bulk}, obtained from the distributional approach.

\subsubsection*{Second junction condition: the angular components}

The angular, or dilatonic, second condition~\eqref{eq:junctionij} takes, in geometrical terms, the form
\begin{equation}\label{eq:second-ang-var-jc}
    \frac{\dd}{\dd \tau} [F_3] - [G_3 \nabla_n \nabla_\tau \varphi] = 2\epsilon\p \frac{\dd A_{D-2}}{\dd \tau} \,.
\end{equation}
Applying the chain rule to the first term and expressing the second as $\nabla_n \nabla_\tau \varphi = \dot{\varphi}_n - K\dot{\varphi}$, the contributions $[G_3\dot\varphi_n]$ cancel between the two pieces. Dividing by $\dot\varphi$, the normal derivative of the dilaton and the extrinsic curvature obey the
jumping relation
\begin{equation}\label{eq:angular-jumps-var}
    [F_{3,\varphi}] + 2\, \Box^{h}\varphi \,[F_{3,Y}] + [G_3 K] = 2 \epsilon \p \frac{\dd A_{D-2}}{\dd \varphi} \,,
\end{equation}
which reproduces~\eqref{eq:angular-jumps} term by term.

\subsubsection*{Mass conservation}

Finally, plugging~\eqref{eq:eom2bij} into the angular second junction condition~\eqref{eq:junctionij} and then, substituting $\Pi_{\tau\tau}$ in terms of the matter part using the other second junction condition, we find the energy balance on the wall,
\begin{equation}
    \frac{\dd m}{\dd \tau} + \epsilon p \frac{\dd{A}_{D-2}}{\dd \tau} - A_{D-2} \,[T_{\tau n}] = 0 \,. \label{eq:second-jc-continuity}
\end{equation}

The distributional and variational approaches therefore agree: both yield the same shell-equation, angular condition and energy balance equation. Moreover, the distributional approach recovers the boundary term and the boundary momenta, which are the basis of the variational route.

\section{Junction conditions for null thin shells} \label{sec:null}

For a null wall, the induced metric on $\Sigma$ degenerates. Although boundary terms for null hypersurfaces are known in Lovelock gravities~\cite{Chakraborty:2018dvi}, their generalization to QT gravities is not available, so we derive the null junction conditions only distributionally.

Let $\Sigma$ be a null hypersurface with coordinates $y^A=(\lambda,\theta^i)$, the same on both sides, where $\lambda$ runs along the null generators and $\theta^i$ are the angles. The tangent vectors are $e^a_{\pm A}=\partial x^a_\pm/\partial y^A$: the one along $\lambda$ is the generator itself and is null, and the ones along $\theta^i$ are spacelike. In the equations below we omit the $\pm$ labels, since they hold on either side:
\begin{equation} \label{eq:knull}
    k^a = \frac{\partial x^a}{\partial \lambda} = e^a_\lambda \,, \qquad k_a k^a = 0 = k_a e^a_i \,.
\end{equation}
Unlike $n^a$ in the non-null case, which lives on the two-dimensional submanifold and can be normalized to $\pm1$, the null generator $k^a$ admits no such normalization and is naturally treated as a $D$-dimensional object. We therefore keep Latin indices for its contractions with $\mathcal{E}_{ab}$ throughout this section, even though spherical symmetry confines $k^a$ to the two-dimensional block, so that $\mathcal{E}_{ab}k^ak^b=\mathcal{E}_{\mu\nu}k^\mu k^\nu$ trivially.

By~\eqref{eq:knull} the induced metric has no component along $\lambda$: it reduces to the angular block $\omega_{ij} = g_{ab}e^a_ie^b_j$ and is therefore \emph{degenerate}, with $k^a$ as its null direction. In spherical symmetry $\omega_{ij}=\varphi^2\Omega_{ij}$, where $\Omega_{ij}$ is the metric of the unit $(D-2)$-sphere, $\dd\Omega^2_{D-2}=\Omega_{ij}\,\dd\theta^i\dd\theta^j$, so the matching $[\omega_{ij}]=0$ becomes $[\varphi]=0$.

Since a degenerate metric cannot raise and lower indices, the basis is completed with an auxiliary null vector,
\begin{equation}
    N_a N^a = 0 \,,\qquad N_a k^a = -1 \,,\qquad N_a e^a_i = 0 \,,
\end{equation}
which gives the completeness relation
\begin{equation}
    g^{ab} = -k^a N^b - N^a k^b + \omega^{ij} e^a_i e^b_j \,.
\end{equation}

Since $k^a$ is both tangent and normal to $\Sigma$, the orthogonality criterion of the non-null case no longer selects a congruence. We therefore take an arbitrary congruence of timelike geodesics crossing the wall, with tangent $u^a$ and proper time $\tau$, such that $\tau=0$ on $\Sigma$, $\tau<0$ in $\mathcal{M}^-$ and $\tau>0$ in $\mathcal{M}^+$, and a displacement along it given by $\dd x^a = u^a \dd\tau$. Smoothness at the wall requires the tangential projections of $u^a$ to match~\cite{Poisson:2009pwt},
\begin{equation}
    \left[-u_a k^a\right] = 0 = \left[u_a e^a_i\right] \,.
\end{equation}

The generator can be written as the gradient of the proper time, 
\begin{equation} \label{eq:k-tau}
    k_a =  -\left(-k_b u^b\right) \partial_a \tau \,,
\end{equation}
where $\tau$ is now regarded as a scalar field near $\Sigma$: each point off the wall lies on a unique geodesic of the congruence, and $\tau$ takes there the value of its proper-time parameter. The wall is the level set $\tau=0$, which is why $\partial_a\tau$ points along $k_a$, and the factor $-k_bu^b$ --- continuous across $\Sigma$ --- follows from $\dd x^a = u^a\dd\tau$. This expression~\eqref{eq:k-tau} replaces $n_a=\epsilon\,\partial_a\ell$: a null vector cannot be normalized, so the normalization is carried by 
\begin{equation}
    \zeta \equiv -(-k_bu^b)^{-1}
\end{equation}
instead. In contrast with $\epsilon$, $\zeta$ is not a sign but a continuous function fixed by the choice of congruence. As before, we write every field as $Q = Q^+ \Theta(\tau) + Q^- \Theta(-\tau)$, so that
\begin{equation}\label{eq:dTheta}
    \partial_a\Theta(\pm\tau) = \pm\, \zeta\, k_a\, \delta(\tau) \,,
\end{equation}
the counterpart of $\partial_a\Theta(\pm\ell)=\pm\epsilon\, n_a\,\delta(\ell)$.

As in the non-null case, we allow matter in the bulk of each region and localized on the wall, so that the stress tensor reads
\begin{equation}\label{eq:tab-null}
    T^{ab} = T^{ab}_+\,\Theta(\tau) + T^{ab}_-\,\Theta(-\tau) - \zeta\, S^{ab}\, \delta(\tau) \,,
\end{equation}
with $S^{ab}$ tangent to $\Sigma$. The factor $-\zeta$ accompanying the surface term is what makes $S^{ab}$ independent of the congruence: the quantities measured by the observers $u^a$ are $-\zeta S^{ab}$, not $S^{ab}$ itself~\cite{Poisson:2009pwt}. The surface stress tensor is
\begin{equation}\label{eq:Sab-null}
    S^{ab} = \mu k^a k^b + j^i \big(k^a e^b_i + e^a_i k^b\big) + p\omega^{ij} e^a_i e^b_j \,,
\end{equation}
with $j^i=0$ by spherical symmetry. Unlike $\sigma$ in the non-null case, $\mu$ is not a proper mass density: a null shell has no rest frame, and both $\mu$ and $p$ depend on the normalization of $k^a$. Throughout this section we make use of $A_{D-2}$, $m$, $\m$ and $\p$ as defined in~\eqref{eq:m-A-def} with $\sigma\to\mu$.

\subsubsection*{First junction condition}

As in the non-null case, a surviving $\delta$ in $\partial_c g_{ab}$ would put one in the connection and $\delta\times\delta$ in the curvature. Removing it requires $[g_{ab}]=0$, whose intrinsic content is the matching $[\omega_{ij}]=0$ already imposed above, that is, $[\varphi]=0$. Continuity of the coordinates then gives $[k^a]=[N^a]=[u^a]=0$.

The metric and the dilaton are continuous, and so are their tangential derivatives: $[\partial_a g_{bc}]k^a = [\partial_a\varphi]k^a = 0$. Hence, the surviving discontinuities are transverse: they point along $k_a$ and their coefficient is obtained by contracting with $N^a$,
\begin{align}
    [\partial_a \varphi] &= -[\varphi_N]\, k_a \,,
    & [\varphi_N] &= N^a\,[\partial_a\varphi] \,, \label{eq:W}\\
    [\partial_c g_{ab}] &= -\chi_{ab}\, k_c \,,
    & \chi_{ab} &= N^c\,[\partial_c g_{ab}] \,, \label{eq:chi}
\end{align}
with $\varphi_N = N^a\partial_a\varphi$ playing the role of $\varphi_n$ and $\chi_{ab}$ that of $[K_{ab}]$. In fact, the object that replaces the extrinsic curvature is the \textit{transverse curvature}
\begin{equation} \label{eq:Cab}
    C_{AB} = \frac12 \left(\pounds_{N} g\right)_{ab}  e^a_A e^b_B \,,
\end{equation}
built with $N^a$ rather than $k^a$, since $K_{AB} = \tfrac12(\pounds_k g)_{ab} e^a_A e^b_B$ is necessarily continuous for a null wall: $k^a$ being tangent to $\Sigma$, $\pounds_k g$ is a tangential derivative. Its jump is fixed by~\eqref{eq:chi},
\begin{equation}
    [C_{AB}] = \frac12\, \chi_{ab}\, e^a_A\, e^b_B \,,
\end{equation}
so that
$\chi_{ab}$ and $[C_{AB}]$ encode the same information. In particular, contracting with $e^a_\lambda = k^a$ gives the component along the generators
\begin{equation}\label{eq:Clala}
    [C_{\lambda\lambda}] = \frac12\, \chi_{ab}\, k^a k^b \,,
\end{equation}
which is the jump of the non-affinity of $k^a$.

\subsubsection*{Singular parts}

Proceeding as in the non-null case, we insert the distributional form of $\varphi$ and $g_{ab}$ into the field equations and collect the terms carrying a $\delta$. The Hessian of the dilaton follows from~\eqref{eq:W} and~\eqref{eq:dTheta}, the Christoffel symbols being bounded,
\begin{equation}\label{eq:hess-null}
    \nabla_a\nabla_b\varphi\big|_{\rm sing.}
    = -\zeta k_a k_b \, [\varphi_N] \, \delta(\tau) \,,
\end{equation}
the analogue of $[\varphi_n]\,n_an_b\,\delta(\ell)$. The singular part of the curvature, in turn, is built from $\chi_{ab}$ and $k^a$, so that the Ricci scalar carries the contraction $\chi_{ab}k^ak^b$ rather than a trace of the extrinsic curvature. The three terms entering
\eqref{eq:einstein-angular} are then 
\begin{align}
    \Box\varphi\big|_{\rm sing.} & = 0 \,, \label{eq:sing-box-null} \\
    R\big|_{\rm sing.} &= - 2 \zeta \,[C_{\lambda\lambda}]\,\delta(\tau) \,, \label{eq:sing-R-null} \\
    \nabla_{[a} \nabla^{b} \varphi \nabla_{b]} \nabla^{a} \varphi |_{\rm sing.} &= -\zeta k^a k^b \nabla_a \nabla_b \varphi|_{\rm reg.} \, [\varphi_N] \,\delta(\tau) \,. \label{eq:sing-boxbox-null}
\end{align}
All of these differ structurally from the non-null case. First,~\eqref{eq:sing-box-null} vanishes because the trace of~\eqref{eq:hess-null} brings a factor $k\cdot k=0$, whereas the same term gave $\epsilon\, [\varphi_n] \,\delta(\ell)$ before; the contribution proportional to $G_{3,\varphi}$ is therefore absent from the angular equation. Second, the two pieces of the singular Ricci scalar --- the double contraction of the jump with the normal and its trace --- exchange roles. In the non-null case the first vanishes, since $g_{\ell\ell}=\epsilon$ is constant along $\ell$, and only the trace survives, giving $-2\epsilon\,[K]$; here the trace is killed by $k\cdot k=0$ and what remains is $[C_{\lambda\lambda}]$, since $k^a$ cannot be normalized. To conclude, in~\eqref{eq:sing-boxbox-null} the piece with the $\delta\times\delta$ term carries $(k\cdot k)^2=0$, so no higher-order distribution survives. In the non-null case the same term was removed by the antisymmetrization; here it is the null character of $k$ that does the job.

\subsubsection*{Second junction condition: the null generator}

We now obtain the second junction conditions relative to the null generator $k^a$. To this end, we insert the singular part~\eqref{eq:hess-null} into~\eqref{eq:einstein-horn},
\begin{equation} \label{eq:Emunu-null}
    \mathcal{E}_{\mu\nu}\big|_{\rm sing.} = \frac{D-2}{2\varphi^{D-2}}
    \zeta \, k_\mu k_\nu \overline{G_3} \, [\varphi_N] \, \delta(\tau) \,,
\end{equation}
where the bar denotes the Vol'pert average, $\varphi_N$ being the only discontinuous quantity across $\Sigma$. The singular part is proportional to $k_a k_b$, so only the $\mu$ term of~\eqref{eq:Sab-null} contributes on the right-hand side, and the factors of $\zeta$ cancel:
\begin{equation}
     \frac{D-2}{2\varphi^{D-2}} \,\overline{G_3} \, [\varphi_N] \, \delta(\tau) = -8\pi\GN \mu \, \delta(\tau) \,.
\end{equation}
In terms of $\m$, this reads
\begin{equation} \label{eq:g3delta-null}
    - \overline{G_3} \, [\varphi_N] \, \delta(\tau) = 2 \m \, \delta(\tau) \,.
\end{equation}
As in the non-null case, the left-hand side contains the product of a jumping variable with a $\delta$. Note, however, that now
\begin{equation}\label{eq:X-null}
    X = \nabla_a\varphi\nabla^a\varphi = -2\varphi_\lambda\varphi_N \,,
\end{equation}
with $\varphi_\lambda \equiv k^a \partial_a \varphi$ tangential and therefore continuous, so $X$ is linear in the jumping variable --- unlike $X=-\epsilon\dot\varphi^2+\epsilon\varphi_n^2$ of the non-null case, which is quadratic in $\varphi_n$. We evaluate the product in two ways: (i) by the Vol'pert chain rule and (ii) by identifying the potential. For the sake of completeness, Appendix~\ref{app:resummation} recovers the same result by resumming the series with $\Theta^{n-1}\delta=\delta/n$ as a prescription.

\textbf{(i) By the Vol'pert chain rule.}
The average~\eqref{eq:volpert-average} of $G_3$ along the jump of $\varphi_N$ reads
\begin{align}
    \overline{G_3} & = \int_0^1 \dd s \, G_3\big(\varphi,-2\varphi_\lambda (\varphi_N^- + s\,[\varphi_N])\big) \notag \\
    & = \frac{1}{[\varphi_N]}\int_{\varphi_N^-}^{\varphi_N^+}\dd z \, G_3(\varphi, -2\varphi_\lambda z) \,,
\end{align}
the second equality following from the change of variable $z=\varphi_N^- + s\,[\varphi_N]$. Substituting the definition~\eqref{eq:G3-def} of $G_3$, the integral is straightforward thanks to the linearity of $X$ in
$\varphi_N$, and gives
\begin{equation} \label{eq:barG3-null}
    \overline{G_3} = 2\varphi^{D-3} \frac{[h(\psi)]}{[\psi]}  \,.
\end{equation}
Plugging this into the field equation,
\begin{equation} \label{eq:jc-volpert-null-null}
     -\varphi^{D-3} \frac{[h(\psi)]}{[\psi]} [\varphi_N] = \m \,,
\end{equation}
and noting that $[\psi] = 2\varphi^{-2}\varphi_\lambda\,[\varphi_N]$, the \textit{shell-equation} for the null case is given by
\begin{equation} \label{eq:jc-volpert-final-null}
    -\varphi^{D-1}\frac{[h(\psi)]}{2\varphi_\lambda}= \m \,.
\end{equation}
For a non-expanding shell, $\varphi_\lambda=0$, the form~\eqref{eq:jc-volpert-null-null} remains valid with $[h]/[\psi]\to h'(\psi)$.

\textbf{(ii) By identifying the potential.}
We now express the singular part of the two-dimensional field equation~\eqref{eq:g3delta-null} as an exact differential using $[\varphi_N]\, \delta(\tau) = (\partial_\tau \varphi_N)|_{\rm sing.}$,
\begin{align} \label{eq:sing-null-null}
    - G_3 (\partial_\tau \varphi_N)|_{\rm sing.} & = 2\m \, \delta(\tau) \,.
\end{align}
Defining, in analogy with $F_3$ in the non-null case,
\begin{align}
    \mathscr{F}_3 & = \int_0^{\varphi_N}  \dd z \, G_3(\varphi,-2\varphi_\lambda z) \notag \\
    & = \frac{\varphi^{D-1}}{\varphi_\lambda} \Bigg( h\left(\frac{1+2\varphi_\lambda\varphi_N}{\varphi^2} \right) - h\left(\frac{1}{\varphi^2} \right) \Bigg) \,,
\end{align}
which satisfies $\partial\mathscr{F}_3/\partial\varphi_N=G_3$,~\eqref{eq:sing-null-null} becomes
\begin{equation}
    - (\partial_\tau \mathscr{F}_3)|_{\rm sing.} = 2 \m \, \delta(\tau) \,.
\end{equation}
Integrating in $\tau$ over an infinitesimal interval $(-\Delta/2,\Delta/2)$ and taking the limit $\Delta\to0$,
\begin{equation} \label{eq:pot-null-lamlam}
    - [\mathscr{F}_3] = 2 \m \,.
\end{equation}
In this case, instead of $\mathscr{F}_3$, it is more convenient to use as potential the generalized Misner--Sharp mass function~\cite{Misner:1964je,Maeda:2007uu}
\begin{equation} \label{eq:mass-function}
    \mathscr{M}(\varphi,X) = \frac12 \varphi^{D-1} h(\psi) \,,
\end{equation}
in terms of which~\eqref{eq:pot-null-lamlam} reads
\begin{equation} \label{eq:pot-null-lamlam-mass}
    -[\mathscr{M}] = \m \, \varphi_\lambda \,.
\end{equation}

\subsubsection*{Second junction condition: the angular components}

Collecting the singular parts, the generalized Einstein tensor reads
\begin{align}
    \mathcal{E}_{ij}\big|_{\rm sing.} &= \frac{\zeta g_{ij}}{2\varphi^{D-3}}  \Big(\overline{G_3} \, [C_{\lambda\lambda}] \notag \\
    & + 2\,\overline{k^a k^b \nabla_a \nabla_b \varphi|_{\rm reg.} G_{3,X}} \, [\varphi_N] \Big)\delta(\tau) \,.
    \label{eq:Eij-null}
\end{align}
We write
\begin{equation}
    k^a k^b \nabla_a \nabla_b \varphi|_{\rm reg.} = \varphi_{\lambda\lambda} - C_{\lambda\lambda} \varphi_\lambda \,,
\end{equation}
with $\varphi_{\lambda\lambda}\equiv k^a\nabla_a(k^b\nabla_b\varphi)$ and $\varphi_\lambda = k^a\nabla_a\varphi$ continuous across $\Sigma$. 

Thus, $\mathcal{E}_{ij}|_{\rm sing.} = -8\pi\GN\zeta p g_{ij}\,\delta(\tau)$ reads
\begin{align}
    & \Big(\overline{G_3} \, [C_{\lambda\lambda}] + 2\,\overline{(\varphi_{\lambda\lambda} - C_{\lambda\lambda}\varphi_\lambda) G_{3,X}} \, [\varphi_N] \Big)\delta(\tau) \notag
    \\ &= -2\p\frac{\dd A_{D-2}}{\dd \varphi}\,\delta(\tau) \,.
    \label{eq:Eij-null-2}
\end{align}

\textbf{(i) By the Vol'pert chain rule.}
As before, two quantities jump simultaneously --- here $\varphi_N$ and $C_{\lambda\lambda}$. Since the complete singular combination is an exact differential in $(\varphi_N,C_{\lambda\lambda})$, as shown in (ii) below, we apply the vectorial Vol'pert rule, in particular~\eqref{eq:product-vector} to the pair $(G_3, C_{\lambda\lambda})$, yielding
\begin{equation} \label{eq:angular-jumps-null}
    [G_3 C_{\lambda\lambda}] + 2 \varphi_{\lambda\lambda}\overline{G_{3,X}} \, [\varphi_N] = -2\p \frac{\dd A_{D-2}}{\dd \varphi} \,.
\end{equation}

In the GR limit $G_{3,X}=0$ and $G_3=2\varphi^{D-3}$ is continuous, so~\eqref{eq:angular-jumps-null} reduces to
\begin{equation}\label{eq:angular-GR-null}
    [C_{\lambda\lambda}]  = -8\pi\GN p\,,
\end{equation}
the standard relation between the pressure of a null shell and the jump of the transverse curvature~\cite{Barrabes:1991ng,Poisson:2009pwt}. 

\textbf{(ii) By identifying the potential.}
Alternatively, the singular part of the angular equation can be written directly as an exact differential in $(\varphi_N,C_{\lambda\lambda})$. After using
\begin{align}
    [\varphi_N] \, \delta(\tau) &= (\partial_\tau \varphi_N)|_{\rm sing.} \,, \\
    [C_{\lambda\lambda}] \, \delta(\tau) &= (\partial_\tau C_{\lambda\lambda} )|_{\rm sing.}\,, \\
    (\partial_\tau G_3 )|_{\rm sing.} & = -2 G_{3,X}\varphi_\lambda (\partial_\tau \varphi_N)|_{\rm sing.} \,,
\end{align}
and taking into account that $\varphi_\lambda$ and $\varphi_{\lambda\lambda}$ are continuous across $\Sigma$, we find the exact differential
\begin{equation} \label{eq:angular-jumps-null-2}
    -\partial_\tau \left(G_3 \bigg(C_{\lambda\lambda} - \frac{\varphi_{\lambda\lambda}}{\varphi_\lambda}\bigg) \right)\bigg|_{\rm sing.} = 2\p \frac{\dd A_{D-2}}{\dd \varphi} \, \delta(\tau) \,.
\end{equation}
Therefore, the potential reads
\begin{equation}\label{eq:pot-null-varphi}
    \Phi_{\varphi}(\varphi_N,C_{\lambda\lambda}) = G_3(\varphi,X)\bigg(C_{\lambda\lambda} - \frac{\varphi_{\lambda\lambda}}{\varphi_\lambda}\bigg) \,.
\end{equation}
Integrating then in $\tau$ over an infinitesimal interval across the shell, only the singular terms contribute, and~\eqref{eq:angular-jumps-null} is simply
\begin{equation}\label{eq:pot-null-varphi-jump}
    -[\Phi_\varphi] = 2\p\frac{\dd A_{D-2}}{\dd\varphi} \,.
\end{equation}
This potential is defined up to a function continuous across the wall.

\subsubsection*{Mass conservation}
We now compute the energy balance on the shell through the mass function $\mathscr{M}$ defined in~\eqref{eq:mass-function}, whose derivative reads
\begin{equation}
    \nabla_\mu \mathscr{M} = \frac12 G_2 \nabla_\mu \varphi - \frac14 G_3\nabla_\mu X \,.
\end{equation}
Taking the two-dimensional field equations~\eqref{eq:einstein-horn}, we find that the following relation holds
\begin{equation}
    \nabla_\mu \mathscr{M} = \frac{8\pi\GN}{(D-2)\Omega_{D-2}}A_{D-2} (T_{\mu\nu} - T g_{\mu\nu})\nabla^{\nu}\varphi \,,
\end{equation}
with $T$ the two-dimensional trace. Differentiating the second junction condition~\eqref{eq:pot-null-lamlam-mass} along the generators, the energy balance on the shell yields
\begin{equation}
    \frac{\dd (m \, \varphi_\lambda)}{\dd \lambda} - A_{D-2} \, [T_{\lambda \lambda} \varphi_N] +  A_{D-2}\varphi_\lambda \, [T_{\lambda N}] = 0 \,,
\end{equation}
where $T_{\lambda\lambda} = k^a k^b T_{ab}$ and $T_{\lambda N} = k^a N^b T_{ab}$. Note that it is independent of the congruence $u^a$ and invariant under a reparametrization $k^a\to \alpha\, k^a$.

For $\lambda=-\varphi$, so that $\varphi_\lambda=-1$, the mass $m$ is therefore conserved whenever $[T_{\lambda \lambda} \varphi_N]$ and $[T_{\lambda N}]$ vanish, in particular in vacuum on both sides.

\subsubsection*{Potentials and the null boundary term}

Although a boundary term for null walls in QT gravities remains unknown, the potentials found above could encode this information. In the non-null case, the potentials $\Phi_{\tau\tau}$ and $\Phi_\varphi$ are, up to normalization, the boundary momenta of the variational problem. By analogy, the mass function~\eqref{eq:mass-function} and the angular potential~\eqref{eq:pot-null-varphi} are the natural candidates for the momenta of a null boundary term, whose jumps across the wall are given by~\eqref{eq:pot-null-lamlam-mass} and~\eqref{eq:pot-null-varphi-jump}.

\section{Discussion} \label{sec:discussion}

\begin{table*}[t]
    \centering
    \begin{tabular}{@{} l l l @{}}
    Case & & Shell-equation \\
    \toprule
    Non-null & & \eqstrut$\displaystyle \frac{\mathsf{m}}{R^{D-3}} = \int^{R_n^-}_{R_n^+} \dd z \, h'\!\left(\frac{1 + \epsilon\dot{R}^2 - \epsilon z^2}{R^2}\right)$ \\
    & & \eqstrut$\displaystyle \phantom{\frac{\mathsf{m}}{R^{D-3}}} \stackrel{\rm trap}{\approx} - \frac{h'(\psi_+) + h'(\psi_-)}{2}\, [R_n]$
    \\
    \midrule
    Null & & \eqstrut$\displaystyle \frac{\mathsf{m}}{R^{D-3}} = \int_{R_N^+}^{R_N^-} \dd z \, h'\!\left(\frac{1+2R_\lambda z}{R^2}\right) = -\frac{R^2}{2R_\lambda}[h(\psi)]$ \\
    & & \eqstrut$\displaystyle \phantom{\frac{\mathsf{m}}{R^{D-3}}} \stackrel{\rm trap}{\approx} - \frac{h'(\psi_+) + h'(\psi_-)}{2}\, [R_N]$
    \\
    \bottomrule
    \end{tabular}
    \caption{Summary of the shell-equations obtained for thin shells, together with their trapezoidal approximations.}
    \label{table:results}
\end{table*}

In this work we have demonstrated the existence of thin shells in Lovelock and QT gravities, which had recently been called into question in~\cite{DiFilippo:2026jpv}. To this end, we have derived the junction conditions in matter-filled QT backgrounds for perfect-fluid thin shells of all types: timelike, spacelike and null. For the non-null ones, we have proved the equivalence between the distributional and variational approaches, that is, the former by studying the delta terms in the field equations and the latter by studying the well-posedness of the problem taking into account the boundary term. As expected, both approaches yield the same results: the shell-equation, the energy balance equation and the relation between the two discontinuous quantities, \ie the jumps of the extrinsic curvature $K$ and of the normal derivative of the dilaton $\varphi_n$. What is more, we found the boundary term from the distributional approach as the singular part of the Lagrangian.

The main novelty of this paper is the treatment of the $\Theta\delta$ products that appear in the field equations when the distributional approach is taken. To give these quantities a definite value we introduce the \textit{Vol'pert chain rule} from the bounded-variation mathematical framework. Since the complete singular combinations appearing in the field equations are exact differentials in the jumping variables, this rule ensures that these singular combinations are well-defined zeroth-order distributions, each given by the jump of a potential. When a single quantity jumps, the discontinuous part multiplying the Dirac delta is simply substituted by its Vol'pert average, which averages the quantity along the segment joining its two one-sided values. In this way, the second junction conditions are straightforward, clean, and valid for any characteristic function $h(\psi)$.

In Table~\ref{table:results} we gather our results for thin shells in Lovelock and QT backgrounds, together with their trapezoidal approximation, obtained by setting $\varphi=r$ in the two-dimensional reduction and denoting by $R$ the location of the shell.

From our results it follows that the thin shell collapse studied in~\cite{Bueno:2024zsx} corresponds to the timelike shell in a vacuum spacetime. In addition, for a null thin shell in vacuum with generators parametrized by $\lambda=-r$ (which corresponds to an ingoing shell), the junction condition~\eqref{eq:pot-null-lamlam-mass}  gives $m=M_+-M_-$, which is constant along the shell for any characteristic function $h(\psi)$.

In contrast, in~\cite{DiFilippo:2026jpv} the null second junction conditions are obtained using $\Theta^{n-1}\delta =\delta/n$ as a prescription. This rule can be derived from Vol'pert theory itself, understanding the Heaviside functions on the left-hand side as their Vol'pert average. However, in that paper, the powers of $\Theta$ are collapsed before this rule is applied, yielding a trapezoidal approximation of the exact integral result --- see Table~\ref{table:results}. Indeed, setting $G_3 = G_3^+\Theta(\tau) + G_3^-\Theta(-\tau)$ directly in~\eqref{eq:g3delta-null}, instead of preserving the powers of $\Theta(\tau)$ inside $G_3(\varphi,X)$ prior to the substitution, reproduces exactly the trapezoidal approximation in Table~\ref{table:results}, which can give a radius-dependent shell mass for theories beyond Einstein--Gauss--Bonnet. The apparent obstruction to thin-shell solutions therefore arises from the difference between the endpoint expression and the exact jump integral.

Regarding mass inflation, our results establish that individual shells are consistent in these theories, as assumed in~\cite{Frolov:2026rcm}. Their crossing requires further analysis that we may investigate in the future.

The construction of a boundary term for null hypersurfaces in QT gravities also remains open. The potentials we found for the null junction conditions are the natural candidates for the momenta of a null boundary term of the variational problem.

Another point worth discussing is the second-order character of the equations of QT gravities in spherical symmetry. This is what ensures that no derivatives of the Dirac delta appear in the generalized field equations. Going beyond the spherically symmetric sector is out of the scope of this work, and there double layers related to higher-order distributions may appear. We also leave this for future work.

\section*{Acknowledgements}

We thank Anthropic's Claude and OpenAI's GPT LLMs for useful discussions.
PB and AVC are supported by the grant PID2022-136224NB-C22, funded by MCIN/AEI/10.13039/501100011033/FEDER, UE. The work of PAC is supported by a Ramón y Cajal fellowship (RYC2023-044375-I) and by a Proyecto de generaci\'on
de conocimiento (PID2024-155685NB-C22) from Spain’s Ministry of Science, Innovation and
Universities. RAH is supported by a Willmore Fellowship at Durham University. ÁJM is supported by the European Union’s Horizon Europe research and innovation programme under the Marie Sklodowska-Curie grant agreement No. 101202045 (RoBHin-ETG). AVC is further supported by a PhD fellowship (FPU24/00899) from Spain's Ministry of Science, Innovation and Universities.

\appendix

\section{Shell-equations by resumming the series} \label{app:resummation}

In this Appendix we rederive the junction conditions for the two-dimensional sector being agnostic of Vol'pert's theory, except for the relation $\Theta^{n-1}\delta=\delta/n$, which can be obtained by other methods and we take as a rule. So, we start from the singular parts of the field equations without the average in both, non-null and null, cases.

\subsection{Non-null case}

Expressing~\eqref{eq:Emunusing2} in terms of the characteristic function~\eqref{eq:G3-def}, it reads
\begin{equation}\label{eq:eom-thetadeltah}
    - h'(\psi) \,[\varphi_n]\, \delta(\ell) = \frac{\m}{\varphi^{D-3}} \, \delta(\ell) \,.
\end{equation}
The quantity that must be given a value is the left-hand side. To do this, note that the jumping variable $\varphi_n=\varphi_n^-+[\varphi_n]\,\Theta(\ell)$ enters $\psi$ quadratically,
\begin{equation} \label{eq:psi-jumps}
    \psi = \frac{1 + \epsilon\dot{\varphi}^2 - \epsilon \varphi_n^2}{\varphi^2}
    = a + b\,\Theta(\ell) + c\,\Theta(\ell)^2 \,,
\end{equation}
for coefficients $a,b,c$ built from $\varphi$, $\dot\varphi$ and $\varphi_n^\pm$. Since this multiplies a $\delta$, the powers of $\Theta$ must be kept, and~\eqref{eq:thetadeltarule} applies to each term of the multinomial expansion,
\begin{align}
    \psi^{n-1}\delta(\ell) & = \sum_{i+j+k=n-1} \binom{n-1}{i,j,k} a^i b^j c^k \, \Theta(\ell)^{j+2k} \, \delta(\ell) \notag \\
    & = \sum_{i+j+k=n-1} \binom{n-1}{i,j,k} \frac{a^i b^j c^k}{1+j+2k} \,\delta(\ell) \notag \\
    & = \delta(\ell) \sum_{i+j+k=n-1} \binom{n-1}{i,j,k} a^i b^j c^k \int_0^1 \dd s\,  s^{j+2k}  \notag \\
    & = \delta(\ell) \int_0^1 \dd s \, (a+bs+cs^2)^{n-1} \notag \\
    & = \delta(\ell) \int_0^1 \dd s \, \psi^{n-1}(s) \,, \label{eq:psindelta} 
\end{align}
since $\int_0^1 \dd s\, s^{j+2k} = 1/(1+j+2k)$ term by term, and where $\psi(s) \equiv a+bs+cs^2$ has been defined in analogy with~\eqref{eq:psi-jumps}, using the continuous variable $s$ instead of $\Theta(\ell)$. Inserting~\eqref{eq:psindelta} into the series expansion of $h'(\psi)$ that follows from its definition~\eqref{eq:hpsi}, and summing over the tower,
\begin{align}
        & h'(\psi)\,\delta(\ell) = \Big(1 + \sum_{n\geq 2} n \alpha_n \psi^{n-1} \Big) \delta(\ell) \notag \\
        & = \int_0^1 \dd s \, \Big(1 + \sum_{n\geq 2} n \alpha_n \psi^{n-1}(s) \Big) \delta(\ell) \notag \\
        & = \int_0^1 \dd s \, h'(\psi(s)=a+bs+cs^2) \delta(\ell) \,.
\end{align}
Therefore, writing $\psi(s)$ back in terms of $\varphi$, $\dot\varphi$ and $\varphi_n^\pm$, the dummy variable $s$ enters only where $\Theta$ did, hence
\begin{align} \label{eq:hprimedelta}
        & h'(\psi)\delta(\ell) = \int_0^1 \!\dd s \, h'\!\left(\frac{1+\epsilon\dot{\varphi}^2-\epsilon (\varphi_n^- + s \, [\varphi_n])^2}{\varphi^2}\right) \!\delta(\ell) \notag \\
        & = \frac{1}{[\varphi_n]} \int_{\varphi_n^-}^{\varphi_n^+} \dd z \, h'\left(\frac{1+\epsilon\dot{\varphi}^2-\epsilon z^2}{\varphi^2}\right) \delta(\ell) \,,
\end{align}
where the change of variable $z = \varphi_n^- + s\,[\varphi_n]$ has been applied in the last step.

Plugging~\eqref{eq:hprimedelta} into the field equations, we obtain
\begin{equation}\label{eq:shell-dist}
    \int_{\varphi_n^{+}}^{\varphi_n^{-}} \dd z \,   h'\! \left( \frac{1 + \epsilon\dot\varphi^2 - \epsilon z^2}{\varphi^2} \right) = \frac{\m}{\varphi^{D-3}} \,,
\end{equation}
which coincides with the junction condition presented in the main text in~\eqref{eq:2jc-volpert-bulk}.

\subsection{Null case}

Expressing now~\eqref{eq:g3delta-null} without the average in terms of the characteristic function~\eqref{eq:G3-def}, it reads
\begin{equation}\label{eq:eom-thetadeltah-null}
    - h'(\psi) \,[\varphi_N]\, \delta(\tau) = \frac{\m}{\varphi^{D-3}} \, \delta(\tau) \,.
\end{equation}

In the null case, the jumping variable is $\varphi_N=\varphi_N^-+[\varphi_N]\,\Theta(\tau)$, so that~\eqref{eq:X-null} and~\eqref{eq:psiXdef} give
\begin{equation}
    \psi = \frac{1+2\varphi_\lambda\varphi_N}{\varphi^2} = \psi_- + [\psi] \, \Theta(\tau) \,.
\end{equation}
Thus, we apply the rule $\Theta^{n-1}\delta=\delta/n$ to each term of the series expansion of the characteristic function
\begin{align}
    \psi^{n-1}\delta(\tau) & =(\psi_-+[\psi]\, \Theta(\tau))^{n-1} \delta(\tau) \notag \\
    & = \sum_{j=0}^{n-1} \binom{n-1}{j}\psi_-^{n-1-j}\,[\psi]^j \frac{1}{j+1} \delta(\tau) \notag \\
    & = \delta(\tau) \sum_{j=0}^{n-1} \frac{1}{n}\binom{n}{j+1}\psi_-^{n-1-j}\,[\psi]^j \notag \\
    & = \delta(\tau) \sum_{k=1}^{n} \frac{1}{n}\binom{n}{k}\psi_-^{n-k}\,[\psi]^{k-1} \notag \\
    & = \frac{1}{n} \delta(\tau) \frac{(\psi_-+[\psi])^{n} - \psi_-^{n}}{[\psi]} \notag \\
    & = \frac{1}{n} \delta(\tau) \frac{[\psi^n]}{[\psi]} \,, \label{eq:psi-term-null}
\end{align}
where we have used $\binom{n-1}{j}\frac{1}{j+1} = \frac{1}{n}\binom{n}{j+1}$.

Hence
\begin{align}
    h'(\psi) \delta(\tau) & = \Bigg(1+\sum_{n\geq2} n\alpha_n \frac{1}{n}\frac{[\psi^n]}{[\psi]} \Bigg)\delta(\tau) \notag \\
    & = \frac{1}{[\psi]}\Bigg([\psi]+\sum_{n\geq2} \alpha_n \, [\psi^n] \Bigg)\delta(\tau) \notag \\
    & = \frac{[h(\psi)]}{[\psi]} \, \delta(\tau) \label{eq:averaged-h-null} \,.
\end{align}
Finally, introducing this expression into~\eqref{eq:eom-thetadeltah-null} one obtains the same junction condition as with the Vol'pert average, namely~\eqref{eq:jc-volpert-null-null}.

\section{Junction conditions for general two-dimensional Horndeski theories}
\label{app:general-junctions}

The Quasi-topological theories considered in the main text comprise a subset of the most general two-dimensional Horndeski theories. The more general class of theories has been used extensively as dynamical models for singularity resolution~\cite{Kunstatter:2015vxa, Colleaux:2019ckh, Carballo-Rubio:2025ntd, Bueno:2025zaj,  Borissova:2026wmn, Borissova:2026krh}. It will therefore be useful to extend the results of the main text to obtain junction conditions for this broader class of theories. We present the non-null case followed by the null case. Our exposition is more compressed than in the main text, as the same techniques are used. 

\subsection{Field equations}

We consider the action~\eqref{eq:2daction} with Lagrangian~\eqref{eq:horn}, but now with $G_2$, $G_3$ and $G_4$ arbitrary functions of $(\varphi,X)$, \ie without imposing~\eqref{eq:Gidentities}. The field equations depend on the $G_w$ only through the specific combinations
\begin{align}
    P(\varphi, X) &\equiv G_2 + X \partial_\varphi \left( G_3 - 2 G_{4, \varphi} \right) \,,
    \\
    Q(\varphi, X) &\equiv 2 G_{4, \varphi} - 2 X \partial_X \left(G_3 - 2 G_{4, \varphi} \right) \,,
    \\
    W(\varphi, X) &\equiv P_{,X} + \frac{1}{2} Q_{,\varphi} \,,
\end{align}
which are $\alpha$, $-2\beta$ and $\omega$ in the notation of Ref.~\cite{Borissova:2026krh}. The field equations read
\begin{align}
\mathcal{E}_{\mu\nu}
&= \frac{D-2}{\varphi^{D-2}}
   \Bigl(
      \frac{Q}{2}
      \bigl(g_{\mu\nu}\,\Box\varphi
            - \nabla_\mu\nabla_\nu\varphi\bigr)
      \notag\\
&\qquad
      - \frac{P}{2}g_{\mu\nu}
      + W\,\nabla_\mu\varphi\,\nabla_\nu\varphi
   \Bigr) \,,
   \label{app:gh-Emunu}\\[1ex]
\mathcal{E}_{ij}
&= \frac{g_{ij}}{2\varphi^{D-3}}
   \Bigl(
      (Q_{,\varphi}+2W)\Box\varphi
      - P_{,\varphi}
      - \frac{Q}{2}R
      \notag\\
&\qquad
      + Q_{,X}
      \bigl[
         (\Box\varphi)^2
         - (\nabla_\mu\nabla_\nu\varphi)
           (\nabla^\mu\nabla^\nu\varphi)
      \bigr]
      \notag\\
&\qquad
      + 2XW_{,\varphi}
      + 2W_{,X}\,\nabla^\mu\varphi\,\nabla_\mu X
   \Bigr) \,,
   \label{app:gh-Eij}
\end{align}
which still satisfy the Bianchi identity~\eqref{eq:bianchirel}. For QT theories,~\eqref{eq:Gidentities} gives
\be
    P^{\rm QT} = G_2 \,, \qquad Q^{\rm QT} = G_3 \,, \qquad W^{\rm QT} = 0 \,, \label{app:gh-QTlimit}
\ee
and~\eqref{app:gh-Emunu}--\eqref{app:gh-Eij} reduce to~\eqref{eq:einstein-horn}--\eqref{eq:einstein-angular}. The $W$-dependent terms therefore contain consequential new information. 

\subsection{Non-null shells}

We use the conventions of Sec.~\ref{sec:nonnull}, with $h_{\tau\tau}=-\epsilon$, $Y=-\epsilon\dot\varphi^2$ and $X = Y + \epsilon\varphi_n^2$. The only singular contribution not already present in the main text is
\be 
    \nabla^\mu\varphi\nabla_\mu X\big|_{\rm sing.} = 2\varphi_n^2\,[\varphi_n]\,\delta(\ell) \,, 
    \label{app:gh-gradX-nonnull}
\ee
which follows from~\eqref{eq:hess-dilaton}. The $W\nabla_\mu\varphi\nabla_\nu\varphi$ term in~\eqref{app:gh-Emunu} is bounded, so the singular part of $\mathcal{E}_{\mu\nu}$ is~\eqref{eq:Emunusing} with $G_3\to Q$. Proceeding as in Sec.~\ref{sec:nonnull}, the shell-equation becomes
\be 
-[F_Q] = 2\m \,, \qquad F_Q \equiv \int_0^{\varphi_n}\dd z\, Q\big(\varphi, Y+\epsilon z^2\big) \,, \label{app:gh-nonnull-shell}
\ee
the analogue of~\eqref{eq:jumpf3}, with $F_Q$ defined as in~\eqref{eq:defF}.

In the angular equation, singular terms featuring $W$ correspond to $2W\,\Box\varphi$ and $2W_{,X}\nabla^\mu\varphi\nabla_\mu X$. By~\eqref{eq:sing-box} and~\eqref{app:gh-gradX-nonnull}, at fixed $(\varphi,Y)$ their singular parts combine into the exact differential $2\epsilon\,\partial_\ell(\varphi_n W)|_{\rm sing.}$, since $\partial(\varphi_nW)/\partial\varphi_n = W + 2\epsilon\varphi_n^2W_{,X}$. The remaining terms are those of the main text with $G_3\to Q$, so the angular junction condition retains the potential form~\eqref{eq:pot-nonull-varphi-jump},
\begin{align}
    -[\Phi_\varphi] &= 2\p\frac{\dd A_{D-2}}{\dd\varphi} \,, \label{app:gh-nonnull-angular}\\
    \Phi_\varphi &= -\epsilon\big(QK + F_{Q,\varphi} + 2 \, \Box^h\varphi\,F_{Q,Y} + 2\varphi_n W\big) \,. \label{app:gh-nonnull-potential}
\end{align}
Compared with~\eqref{eq:angular-jumps}, the only changes are $G_3\to Q$ and the new term $2\,[\varphi_nW]$. The energy conservation~\eqref{eq:second-jc-continuity-distrib} requires only $\nabla_aT^{ab}=0$ and is therefore unchanged.

\subsection{Null shells}

We use the null frame and conventions of Sec.~\ref{sec:null}, in which $X=-2\varphi_\lambda\varphi_N$ by~\eqref{eq:X-null}, $\varphi_\lambda$ and $\varphi_{\lambda\lambda}$ are continuous, and $\varphi_N$ and $C_{\lambda\lambda}$ may jump across the shell. In addition to~\eqref{eq:hess-null}--\eqref{eq:sing-boxbox-null}, we need
\be 
\nabla^\mu\varphi\nabla_\mu X\big|_{\rm sing.} = -2\zeta\varphi_\lambda^2\,[\varphi_N]\,\delta(\tau) \,. \label{app:gh-gradX-null}
\ee
As in the main text, the piece of $\nabla_{[\mu}\nabla^\nu\varphi\nabla_{\nu]}\nabla^\mu\varphi$ quadratic in $\delta$ has coefficient $(k\cdot k)^2=0$ and hence vanishes.

To obtain the equation of motion of the shell, we again note that the term containing $W$ in~\eqref{app:gh-Emunu} is bounded, and the singular part of $\mathcal{E}_{\mu\nu}$ is therefore~\eqref{eq:Emunu-null} with $G_3\to Q$. The analogue of~\eqref{eq:pot-null-lamlam} is then
\be 
    -[\mathscr{F}_Q] = 2\m \,, \qquad \mathscr{F}_Q \equiv \int_0^{\varphi_N}\dd z\, Q(\varphi,-2\varphi_\lambda z) \,. \label{app:gh-null-shell}
\ee
This result applies for both expanding and non-expanding, \ie $\varphi_\lambda = 0$, null shells. 

To address the angular equation, we first note that by~\eqref{eq:sing-box-null},~\eqref{eq:sing-R-null},~\eqref{eq:sing-boxbox-null} and~\eqref{app:gh-gradX-null}, the term in square brackets in~\eqref{app:gh-Eij} has a singular part that can be expressed as an exact differential, $\dd \Phi_\varphi$, with
\be 
    \Phi_\varphi = QC_{\lambda\lambda} + 2\varphi_{\lambda\lambda}\int_0^{\varphi_N}\dd z\,Q_{,X}(\varphi,-2\varphi_\lambda z) + 2\varphi_\lambda W \,, \label{app:gh-null-potential}
\ee
and the differential taken with the variables $(\varphi_{N}, C_{\lambda\lambda})$ with fixed $(\varphi, \varphi_\lambda, \varphi_{\lambda\lambda})$. Matching to the source as in Sec.~\ref{sec:null} gives $-[\Phi_\varphi] = 2\p\,\dd A_{D-2}/\dd\varphi$, or explicitly
\begin{align} 
    [QC_{\lambda\lambda}] &+ 2\varphi_{\lambda\lambda}\int_{\varphi_N^-}^{\varphi_N^+}\dd z\,Q_{,X}(\varphi,-2\varphi_\lambda z) \notag\\
    &+ 2\varphi_\lambda\,[W] = -2\p\frac{\dd A_{D-2}}{\dd\varphi} \,, \label{app:gh-null-angular} 
\end{align}
which generalizes~\eqref{eq:angular-jumps-null} by $G_3\to Q$ and the new term $2\varphi_\lambda\, [W]$. For $\varphi_\lambda\neq0$, the integral can be explicitly evaluated, and we have 
\be 
\Phi_\varphi = Q\,(C_{\lambda\lambda}-\varphi_{\lambda\lambda}/\varphi_\lambda) + 2\varphi_\lambda W \,,
\ee 
in agreement with~\eqref{eq:pot-null-varphi} when $W=0$. On the other hand, when the shell is non-expanding with $\varphi_\lambda=\varphi_{\lambda\lambda}=0$, \eqref{app:gh-null-shell} and~\eqref{app:gh-null-angular} reduce to $-Q(\varphi,0)[\varphi_N]=2\m$ and $Q(\varphi,0)[C_{\lambda\lambda}]=-2\p\,\dd A_{D-2}/\dd\varphi$. This means a shell of constant area can still carry a nonvanishing density.


\bibliographystyle{JHEP-2}
\bibliography{Gravities}
\noindent

\end{document}